\documentclass[twocolumn,times]{aastex63}
\usepackage{amsmath}
\usepackage{newtxmath}
\usepackage{mleftright}
\mleftright

\defcitealias{Ireland:2021db}{I}

\newcommand{\appropto}{\mathrel{\vcenter{
  \offinterlineskip\halign{\hfil$##$\cr
    \propto\cr\noalign{\kern2pt}\sim\cr\noalign{\kern-2pt}}}}}
\shorttitle{Magnetic Braking of Accreting T Tauri Stars II}
\shortauthors{Ireland et al.}
\begin{document}

\title{Magnetic Braking of Accreting T Tauri Stars II: Torque Formulation Spanning Spin-Up and Spin-Down Regimes}

\author[0000-0002-8833-1204]{Lewis G. Ireland}
\affil{Department of Physics and Astronomy, University of Exeter, Stocker Road, Exeter, EX4 4QL, UK}

\author[0000-0001-9590-2274]{Sean P. Matt}
\affil{Department of Physics and Astronomy, University of Exeter, Stocker Road, Exeter, EX4 4QL, UK}

\author[0000-0003-0204-8190]{Claudio Zanni}
\affil{INAF - Osservatorio Astrofisico di Torino, Strada Osservatorio 20, {10025 Pino Torinese}, Italy}

% \author[0000-0001-7788-3727]{George Pantolmos}
% \affil{Univ. Grenoble Alpes, CNRS, IPAG, 38000 Grenoble, France}

\correspondingauthor{Lewis G. Ireland}
\email{L.G.Ireland@exeter.ac.uk}



\begin{abstract}

% This example manuscript is intended to serve as a tutorial and template for
% authors to use when writing their own AAS Journal articles. The manuscript
% includes a history of \aastex\ and documents the new features in the
% previous version, 6.0, as well as the new features in version 6.1. This
% manuscript includes many figure and table examples to illustrate these new
% features.  Information on features not explicitly mentioned in the article
% can be viewed in the manuscript comments or more extensive online
% documentation. Authors are welcome replace the text, tables, figures, and
% bibliography with their own and submit the resulting manuscript to the AAS
% Journals peer review system.  The first lesson in the tutorial is to remind
% authors that the AAS Journals, the Astrophysical Journal (ApJ), the
% Astrophysical Journal Letters (ApJL), and Astronomical Journal (AJ), all
% have a 250 word limit for the abstract.  If you exceed this length the
% Editorial office will ask you to shorten it.

The magnetic interaction between a classical T Tauri star and its surrounding accretion disk is thought to influence its rotational evolution.
We use 2.5D magnetohydrodynamic, axisymmetric simulations of star-disk interaction, computed via the PLUTO code, to calculate the net torque acting on these stars. 
We divide the net torque into three contributions: accretion (spin-up), stellar winds (spin-down), and magnetospheric ejections (MEs) (spin-up or down).
In Paper~\citetalias{Ireland:2021db}, we explored interaction regimes in which the stellar magnetosphere truncates the inner disk at a location spinning faster than the star, resulting in a strong net spin-up contribution from accretion and MEs (``steady accretion" regime).
In this paper, we investigate interaction regimes in which the truncation radius gets closer to and even exceeds corotation, where it is possible for the disk material to gain angular momentum and be periodically ejected by the centrifugal barrier (``propeller" regime). 
This reduces the accretion torque, can change the sign of the ME torque, and can result in a net stellar spin-down configuration. These results suggest it is possible to have a net spin-down stellar torque even for truncation radii within the corotation radius ($R_\text{t} \gtrsim 0.7 R_\text{co}$).
We fit semi-analytic functions for the truncation radius, and the torque associated with star-disk interaction (i.e., the sum of accretion and ME torques) and stellar wind, allowing for the prediction of the net stellar torque for a parameter regime covering both net spin-up and spin-down configurations, as well as the possibility of investigating rotational evolution via 1D stellar evolution codes.

\end{abstract}

% \keywords{convection --- magnetohydrodynamics (MHD) --- stars: fundamental parameters --- 
% stars: low-mass --- stars: interiors --- stars: rotation}



\section{Introduction}

Classical T Tauri stars (CTTS) are observed to still be actively accreting material from a circumstellar accretion disk \citep{1994AJ....108.1056E,1998ApJ...492..323G,1998ApJ...495..385H,2014ApJ...790...47I,2017A&A...600A..20A}. Additionally, these pre-main sequence (PMS) solar-like stars are also still undergoing gravitational contraction. 
As a result, it could be expected that these objects are spinning up over time or driven to near-break-up rotation rates over several million years. However, many of these stars appear to have fairly constant spin distributions over this timescale \citep{1993A&A...272..176B,1993AJ....106..372E,2004AJ....127.1029R,2009IAUS..258..363I,2019A&A...632A...6G}, with a majority rotating much lower than break-up \citep[see, e.g.,][]{1993A&A...272..176B,2004AJ....127.1029R,2007prpl.conf..297H}. Therefore, the current theoretical explanation for the rotation evolution of CTTS is incomplete.

Given that CTTS are magnetically active, with $\sim$kG strength multipolar fields \citep[see, e.g.,][]{2007ApJ...664..975J,10.1111/j.1365-2966.2008.13687.x,2014MNRAS.437.3202J,2019MNRAS.483L...1D,2020MNRAS.491.5660D}, the magnetic connection between the star and its disk has been theorized to allow for angular momentum removal from the system, allowing stars that are actively accreting to effectively reach spin-equilibrium, or even spin-down, over time \citep[see, e.g.,][]{1990RvMA....3..234C,1991ApJ...370L..39K,2002ApJ...565.1205U}. 
In the classic \citet{1979ApJ...234..296G} model, angular momentum is extracted from the star by its disk, along closed field lines connecting the stellar surface to the disk beyond the corotation radius, at which the disk rotates more slowly than the star. However, it has been demonstrated analytically, and via magnetohydrodynamic (MHD) simulations of star-disk interaction (SDI), that the spin-down torque efficiency is vastly reduced compared to the \citet{1979ApJ...234..296G} picture: star-disk differential rotation results in field lines ``twisting", which inflates and opens the magnetosphere at mid-latitudes \citep[see][and references therein]{2005MNRAS.356..167M}, and the poloidal field intensity becomes diluted outside the corotation radius \citep{2000MNRAS.317..273A,2009A&A...508.1117Z}.

Another proposed mechanism is magnetized stellar winds, which provide a spin-down torque by extracting angular momentum via open magnetic field lines connected to the stellar surface \citep{2005ApJ...632L.135M}. 
\citet{2008ApJ...681..391M} show a stellar wind mass flux of $\sim 10 \%$ the mass-accretion rate is required for the stellar wind alone to counteract the spin-up accretion torque, which is unlikely achievable if powered solely by the star's rotational, thermal, and magnetic energies; therefore, an extra energy source would be required to drive such outflows \citep{2007IAUS..243..299M}. 
\citet{2005ApJ...632L.135M} proposed accretion-powered winds as a potential mechanism, but there are energetic limitations \citep{2008ApJ...681..391M, 2011ApJ...727L..22Z}, and the possible mechanisms for transferring this power are still under investigation \citep{2008ApJ...689..316C, 2009ApJ...706..824C}.

More recently, an additional ejection mechanism has been recognized as potentially important: the opening and reconnection of field lines, named magnetospheric ejections (MEs), which occurs due to the build up of toroidal field pressure as a result of the star-disk differential rotation \citep{Goodson_1999,refId0}. 
MEs remove angular momentum from the system as a whole, but since they are magnetically connected to both the star and disk, MEs can also transfer angular momentum between the two. The ME torque contribution exerted onto the star can either spin-up or spin-down the star, depending on whether the angular momentum extracted from the disk is located in a region rotating faster or slower than corotation, respectively.

In order to know whether various proposed mechanisms can explain the general slow rotation rates of CTTS and the observed evolution of their spin-rate distributions, several authors have computed the angular momentum evolution during this phase, including torques arising from SDI. \citet{1993A&A...274..309C}, \citet{1994ApJ...428..760Y,1995ApJ...442..768Y}, \citet{1996MNRAS.280..458A}, and \citet{2010ApJ...714..989M} included torques just from accretion and the magnetic connection, and \citet{2012ApJ...745..101M} considered accretion-powered stellar winds. \citet{2019A&A...632A...6G} considered torques including MEs, but their formulation was based only on four simulations, and employs coefficients that appear to be guided by simulations, rather than explicitly fitted.
In Paper \citetalias{Ireland:2021db}, we ran a suite of MHD simulations of accreting, magnetized stars, that included stellar winds and MEs, exploring the parameter regime where the truncation radius was well within corotation, i.e., the ``steady accretion" regime. Fitting to the simulation results, we formulated physical prescriptions to predict the net torque on the star, for a parameter regime that represented stars over a wide range of ages. All of our simulations were in the net spin-up regime (i.e., they had a net positive torque). 
In this paper, we expand on our parameter regime from Paper \citetalias{Ireland:2021db}, in the pursuit of SDI systems that are in a net stellar spin-down configuration. We perform 47 2.5D axisymmetric SDI simulations, changing the initial disk density, the stellar surface magnetic field strength, and the stellar rotation rate, to capture how the net stellar torque (contributed by the stellar wind, accretion, and MEs) depends on both stellar and disk global properties. The numerical method by which we perform these simulations is identical to that of Paper \citetalias{Ireland:2021db}.

Our parameter range reaches into the ``propeller" regime, where the azimuthal velocity of the star's outer magnetosphere exceeds the Keplerian velocity at disk truncation \citep{Romanova_2005}, i.e., where the disk is truncated near and outside of the corotation radius. Observations of the propeller regime have been discussed in relation to various astrophysical systems, such as rapidly-rotating neutron stars, white dwarfs in cataclysmic variables, and CTTS \citep[see, e.g.,][]{1986ApJ...308..669S,1993A&A...269..319T,Cui_1997,Alpar_2001,Ek__2003,Mori_2003}. The propeller regime has been investigated both analytically \citep{10.1093/mnras/186.4.779,1997ASPC..121..241L,Lovelace_1999,ikhsanov_2002,Rappaport_2004,Ek_i_2005}, and via MHD simulations \citep{1985A&A...151..361W,Romanova_2003,Romanova_2004,Romanova_2005,2006ApJ...646..304U,2009MNRAS.399.1802R}.
It has been shown that for a rapidly rotating star with a strong magnetic field, a significant proportion of accreting material is in fact centrifugally expelled as it reaches the inner region of the disk and is redirected as a propeller-driven outflow, allowing the star to rapidly spin down \citep{1973ApJ...179..585D,1975A&A....39..185I,1992ans..book.....L}.
For example, MHD simulations performed by \citet{Romanova_2005} and \citet{refId0} demonstrate the following quasi-periodic behaviour: (1) disk material accumulates at the inner region of the disk, which moves the truncation radius closer to the star; (2) some material can then accrete onto the star, reducing the mass of material at the disk inner edge;  (3) remaining inner-disk material gains angular momentum and is ejected at the centrifugal barrier, which moves the truncation radius further away from the star, for the cycle to repeat. 
In this regime, accretion becomes intermittent, or can even be inhibited completely, and the net effect acts to remove angular momentum from the star. Therefore, it is possible that the propeller mechanism is responsible for the slow rotation rates observed in CTTS.

We present an updated, more general torque formulation that is compatible with our new SDI simulations and those found in Paper~\citetalias{Ireland:2021db}, representing simulations across both the spin-up and spin-down regimes. We use an identical stellar wind torque formulation to that in Paper \citetalias{Ireland:2021db} across our new parameter regime. 
We propose a formulation for the truncation radius and the sum of the accretion and stellar ME torques, introduced as the ``SDI" torque, which depends on the relative locations of the truncation and corotation radii for the system.
In Section \ref{sec:num_method}, we briefly introduce the numerical setup of our simulations, normalizations, and the parameter regime explored. In Section~\ref{sec:sims}, we present a qualitative description of their behavior across three different regimes investigated in this parameter study. We also introduce the global quantities used, such as the mass flow rate, torque, and unsigned magnetic flux, investigating changes in their behavior across different regimes. In Section~\ref{sec:torque_formulation}, we introduce our updated torque formulation for the truncation radius, the SDI torque (accretion + MEs), and the stellar wind, allowing one to calculate the net stellar torque a priori. In Section~\ref{sec:dis_conc}, we discuss and conclude our findings.

\section{Numerical Method}\label{sec:num_method}

\subsection{Numerical Setup and Initial Conditions}\label{sec:setup_initial}

Here, we briefly describe the numerical setup and initial conditions of the simulations in this parameter study. A more detailed identical setup, including boundary conditions adopted and relevant formulae, can be found in Paper \citetalias{Ireland:2021db}. 

For this work, we perform 2.5D axisymmetric MHD simulations (including resistive and viscous effects) with a spherical computational domain, covering [1,50.756521] $R_\star$ in the $R$ direction across 320 logarithmic grid cells ($R_\star$ is the stellar radius), and [0,$\pi$] in the $\theta$ direction across 256 uniform grid cells. The grid cell sizes satisfy $\Delta R \sim R \Delta \theta$. These simulations use our adopted caloric equation of state \citep[see appendix of][]{Pantolmos2020}, where the specific heats and adiabatic index $\gamma$ are temperature-dependent; a hot stellar wind behaves near-isothermally ($\gamma=1.05$), whereas a cold accretion disk behaves adiabatically ($\gamma=5/3$).

We initialize our simulation domain with the following: a stellar corona, using density and pressure profiles from a 1D spherically symmetric, isentropic, transonic Parker wind solution; an initial Keplerian viscous accretion disk, adopting an $\alpha$ viscosity parameterization \citep{1973A&A....24..337S}; and an initial dipolar field configuration.

\subsection{Units and Normalization}\label{sec:units}

Simulations in this paper are performed in dimensionless units, therefore, we list normalization factors required to convert results into physical units. We express length in units of $R_\star$, density in units of the initial stellar surface coronal density, $\rho_\star$, velocities in units of stellar surface Keplerian velocity, $v_{\text{K},\star}=(GM_\star/R_\star)^{1/2}$ ($G$ is the gravitational constant, and $M_\star$ is the stellar mass), time in units of $t_0 = R_\star/v_{\text{K},\star}$, magnetic field strength in units of $B_0 = (4 \pi \rho_\star v_{\text{K},\star}^2)^{1/2}$, mass flux in units of $\dot{M}_0 = \rho_\star R_\star^2 v_{\text{K},\star}$, and torque in units of $\dot{M}_0 = \rho_\star R_\star^3 v_{\text{K},\star}^2$.

By adopting $R_\star=2 \, R_\odot$, $M_\star=0.5 \, M_\odot$, and $\rho_\star = 10^{-12}$ g cm$^{-3}$, we can make direct comparisons with young stars using the following normalizations:

\begin{align}\label{eq:norm}
\begin{gathered}
v_{\text{K},\star}  = 218.38  \left(\tfrac{M_\star}{0.5 \, M_\odot}\right)^{1/2}  \left(\tfrac{R_\star}{2 \, R_\odot}\right)^{-1/2}  \,  \text{km s}^{-1}\\
B_0  = 77.41 \left(\tfrac{\rho_\star}{10^{-12} \, \text{g} \, \text{cm}^{-3}}\right)^{1/2}  \left(\tfrac{M_\star}{0.5 \, M_\odot}\right)^{1/2}  \left(\tfrac{R_\star}{2 \, R_\odot}\right)^{-1/2} \,  \text{G} \\
t_0 =  0.074 \left(\tfrac{M_\star}{0.5 \, M_\odot}\right)^{-1/2} \left(\tfrac{R_\star}{2 \, R_\odot}\right)^{3/2} \, \text{days} \\
\dot{M}_0 =  6.71 \times 10^{-9} \left(\tfrac{\rho_\star}{10^{-12} \, \text{g cm}^{-3}}\right) \left(\tfrac{M_\star}{0.5 \, M_\odot}\right)^{1/2} \left(\tfrac{R_\star}{2 \, R_\odot}\right)^{3/2} \, M_\odot \, \text{yr}^{-1} \\
\dot{J}_0 =  1.29 \times 10^{36} \left(\tfrac{\rho_\star}{10^{-12} \, \text{g cm}^{-3}}\right) \left(\tfrac{M_\star}{0.5 \, M_\odot}\right) \left(\tfrac{R_\star}{2 \, R_\odot}\right)^{2} \, \text{erg}.
\end{gathered}
\end{align}

\subsection{Simulation Parameter Study}\label{sec:sim_params}

Analogous to Paper \citetalias{Ireland:2021db}, we investigate the following parameter space:

\begin{enumerate}
\item{Initial disk density, $\rho_{\text{d,}\star}$;} 
\item{Surface polar magnetic field strength, $B_\star$;\footnote{$B_\star$ is controlled via the input parameter $v_\text{A}/v_\text{esc}$, i.e., the ratio of the surface polar Alfvén velocity $v_\text{A} = B_\star / (4 \pi \rho_\star)^{1/2}$ and the stellar escape velocity $v_\text{esc}=(2GM_\star/R_\star)^{1/2}$.}}
\item{Stellar break-up fraction, $f=\Omega_\star R_\star / v_{\text{K},\star}$;}
\end{enumerate}
where $\Omega_\star$ is the stellar rotation rate. The stellar rotation period can be written as

\begin{equation}\label{eq:period}
P_\star = 7.44 \left(\frac{f}{0.0625}\right)^{-1} \left(\frac{M_\star}{0.5 \, M_\odot}\right)^{-1/2} \left(\frac{R_\star}{2 \, R_\odot}\right)^{3/2} \, \text{days}.
\end{equation}
The viscous and resistive transport coefficients are fixed at $\alpha_\text{v}=0.2$ and $\alpha_\text{m}=0.2$, respectively, the disk thermal aspect-ratio at $\epsilon=0.075$, and the stellar wind sound speed at the stellar surface at $c_{\text{s},\star} = 0.35 v_{\text{K},\star}$.

This study consists of 47 simulations in total. 
Input parameters and outputted variables (in simulation units) for all simulations are listed in Table~\ref{tab:Pluto_sims}, and normalization factors to convert these into physical units are conveniently listed in Table~\ref{tab:conversion}.
Each simulation covers a period that roughly corresponds to 20 stellar rotation periods for a break-up fraction of $f=0.05$ ($t_0=2513.2742$), which typically covers a period on the order of weeks or months (when scaled to T Tauri star parameters), where it can be assumed that changes in the stellar rotation rate are negligible (justifying our use of fixed $f$).
Outputted variables are time-averages over the latter half of each simulation's temporal domain (unless otherwise stated in Table~\ref{tab:Pluto_sims}), where models appear quasi-steady or quasi-periodic, avoiding the transient adjustment from the initial state.
All simulations with $f < 0.05$, and those with $f=0.05$ and $\rho_{\text{d},\star}/\rho_\star \geq 100$, are from the parameter study of Paper \citetalias{Ireland:2021db}. New simulations performed for this paper explore lower disk densities $6.25 \leq \rho_{\text{d},\star}/\rho_\star \leq 50$, an additional magnetic field strength $B_\star/B_0=19.5$, and higher break-up fractions $0.05 \leq f \leq 0.1$.

\begin{deluxetable*}{ccccc|ccccccccccccccc}
\tabletypesize{\footnotesize}

\tablecaption{Variable Input Parameters and Outputted Global Variables for All Simulations\label{tab:Pluto_sims}}
\setlength{\tabcolsep}{3.5pt}
\tablehead{\rule{0pt}{4ex} Model & $f$ & $\frac{\rho_{\text{d,}\star}}{\rho_\star}$ & $\frac{v_\text{A}}{v_\text{esc}}$$^{\text{a}}$ & $\frac{B_\star}{B_0}$$^{\text{b}}$ & State & $\frac{\dot{M}_\text{wind}}{\dot{M}_0}$ & $\frac{\dot{M}_{\text{ME},\star}}{\dot{M}_0}$ & $\frac{\dot{M}_{\text{ME,out}}}{\dot{M}_0}$ & $\frac{\dot{M}_\text{acc}}{\dot{M}_0}$ & $\frac{\dot{J}_\text{wind}}{\dot{J}_0}$ & $\frac{\dot{J}_{\text{ME},\star}}{\dot{J}_0}$ & $\frac{\dot{J}_\text{acc}}{\dot{J}_0}$ & $\frac{\dot{J}_\star}{\dot{J}_0}$ & $\frac{\Phi_\text{wind}}{\Phi_0}$ & $\frac{\langle r_\text{A} \rangle}{R_\star}$ & $\frac{R_\text{t}}{R_\star}$ & $\frac{R_\text{t}}{R_\text{co}}$ & $\Upsilon_\text{wind}$ & $\Upsilon_\text{acc}$ \\
&&&&&&$[10^{-3}]$&$[10^{-3}]$&$[10^{-3}]$&&&&&&&&&&$[10^4]$&}
\startdata
1 & 0.01 & 100.0 & 9.2 & 13 & 1 & 4.25 & -5.8 & 64.1 & -0.375 & 0.0663 & -0.198 & -0.531 & -0.663 & 8.99 & 39.5 & 4.59 & 0.213 & 1.34 & 373 \\
2 & 0.01 & 200.0 & 9.2 & 13 & 1 & 5.25 & -8.69 & 111 & -0.756 & 0.113 & -0.325 & -0.972 & -1.18 & 12.2 & 46.3 & 3.76 & 0.175 & 2.01 & 164 \\
3 & 0.01 & 200.0 & 4.6 & 6.5 & 1 & 5.16 & -32.6 & 133 & -0.803 & 0.11 & -0.403 & -0.829 & -1.12 & 11.5 & 46.1 & 2.34 & 0.109 & 1.81 & 38.6 \\
4 & 0.01 & 300.0 & 9.2 & 13 & 1 & 5.6 & -15.9 & 190 & -1.2 & 0.159 & -0.504 & -1.45 & -1.79 & 15.3 & 53.2 & 3.28 & 0.152 & 2.97 & 102 \\
5 & 0.01 & 400.0 & 9.2 & 13 & 1 & 5.8 & -26.4 & 285 & -1.6 & 0.209 & -0.647 & -1.84 & -2.28 & 17.8 & 59.9 & 3 & 0.139 & 3.87 & 75.7 \\
6 & 0.01 & 200.0 & 18.4 & 26 & 1 & 3.23 & -2.71 & 166 & -0.796 & 0.117 & -0.411 & -1.28 & -1.58 & 13.3 & 60.1 & 5.88 & 0.273 & 3.9 & 650 \\
7 & 0.025 & 100.0 & 9.2 & 13 & 1 & 5.35 & -4.39 & 62.9 & -0.579 & 0.196 & -0.209 & -0.8 & -0.813 & 11.1 & 38.3 & 4.13 & 0.353 & 1.62 & 220 \\
8 & 0.025 & 200.0 & 9.2 & 13 & 1 & 6.12 & -13.2 & 134 & -1.08 & 0.318 & -0.408 & -1.32 & -1.41 & 14.9 & 45.6 & 3.44 & 0.294 & 2.55 & 113 \\
9 & 0.025 & 300.0 & 18.4 & 26 & 2 & 4.78 & -8.9 & 350 & -1.78 & 0.449 & -0.724 & -2.61 & -2.88 & 19.9 & 61.1 & 4.54 & 0.388 & 5.85 & 287 \\
10 & 0.025 & 300.0 & 9.2 & 13 & 1 & 6.48 & -19.8 & 211 & -1.5 & 0.427 & -0.543 & -1.79 & -1.9 & 17.8 & 51.3 & 3.09 & 0.264 & 3.44 & 81.3 \\
11 & 0.025 & 400.0 & 9.2 & 13 & 1 & 6.77 & -35.1 & 301 & -1.91 & 0.545 & -0.688 & -2.21 & -2.36 & 20.4 & 56.6 & 2.78 & 0.238 & 4.33 & 63.9 \\
12 & 0.025 & 400.0 & 6.9 & 9.75 & 1 & 6.7 & -64.7 & 301 & -2 & 0.522 & -0.793 & -2.17 & -2.44 & 20.2 & 55.5 & 2.27 & 0.194 & 4.31 & 34.6 \\
13 & 0.05 & 6.25 & 4.6 & 6.5 & 2 & 6.5 & -0.878 & 6.6 & -0.0833 & 0.119 & -0.00442 & -0.125 & -0.00992 & 6.16 & 19.1 & 4.43 & 0.602 & 0.413 & 988 \\
14 & 0.05 & 6.25 & 9.2 & 13 & 2 & 3.88 & -3.4 & 18.8 & -0.0657 & 0.149 & 0.0642 & -0.146 & 0.0675 & 7.7 & 27.7 & 5.58 & 0.757 & 1.08 & 1.35e+05 \\
15 & 0.05 & 6.25 & 13.8 & 19.5 & 2 & 3.49 & -30.6 & 32.7 & -0.0228 & 0.25 & 0.201 & -0.0193 & 0.431 & 10.5 & 37.8 & 6.62 & 0.898 & 2.24 & 7.67e+05 \\
16 & 0.05 & 6.25 & 18.4 & 26 & 3 & 2.93 & -61.4 & 56.7 & -0.00146 & 0.292 & 0.329 & -0.0066 & 0.615 & 11.9 & 44.5 & 7.6 & 1.03 & 3.43 & 1.05e+07 \\
17 & 0.05 & 12.5 & 13.8 & 19.5 & 2 & 3.64 & -4.46 & 35.6 & -0.136 & 0.264 & 0.166 & -0.291 & 0.14 & 10.9 & 38.1 & 6.37 & 0.864 & 2.32 & 2.75e+05 \\
18 & 0.05 & 12.5 & 18.4 & 26 & 2 & 3.17 & -41.8 & 47.8 & -0.093 & 0.333 & 0.323 & -0.161 & 0.495 & 12.8 & 45.8 & 7.05 & 0.957 & 3.67 & 6.95e+05 \\
19 & 0.05 & 12.5 & 4.6 & 6.5 & 2 & 6.83 & -1.04 & 11.7 & -0.147 & 0.133 & -0.0338 & -0.216 & -0.117 & 6.43 & 19.7 & 4.14 & 0.562 & 0.428 & 239 \\
20 & 0.05 & 25.0 & 13.8 & 19.5 & 2 & 3.97 & -2.48 & 59.4 & -0.305 & 0.297 & 0.0721 & -0.65 & -0.28 & 11.7 & 38.7 & 5.87 & 0.797 & 2.43 & 2.94e+03 \\
21 & 0.05 & 50.0 & 13.8 & 19.5 & 2 & 3.54 & -8.88 & 345 & -0.655 & 0.26 & 0.0829 & -1.27 & -0.924 & 10.9 & 38.2 & 5.09 & 0.691 & 2.36 & 579 \\
22 & 0.05 & 100.0 & 9.2 & 13 & 2 & 7.76 & -6.87 & 104 & -0.948 & 0.551 & -0.243 & -1.28 & -0.972 & 15 & 37.6 & 3.61 & 0.49 & 2.07 & 133 \\
23 & 0.05 & 100.0 & 6.9 & 9.75 & 1 & 8.16 & -10.7 & 76.3 & -0.779 & 0.449 & -0.204 & -0.971 & -0.725 & 13.1 & 33.2 & 3.17 & 0.43 & 1.48 & 88.8 \\
24 & 0.05 & 100.0 & 4.6 & 6.5 & 1 & 8.24 & -16.4 & 60.1 & -0.652 & 0.356 & -0.216 & -0.712 & -0.572 & 11.5 & 29.4 & 2.62 & 0.355 & 1.14 & 47.4 \\
25 & 0.05 & 200.0 & 6.9 & 9.75 & 1 & 8.49 & -27.3 & 144 & -1.33 & 0.668 & -0.4 & -1.5 & -1.23 & 16.8 & 39.7 & 2.64 & 0.358 & 2.34 & 51.7 \\
26 & 0.05 & 200.0 & 9.2 & 13 & 1 & 8.38 & -16.2 & 175 & -1.54 & 0.783 & -0.433 & -1.88 & -1.53 & 18.5 & 43.2 & 3.11 & 0.422 & 2.88 & 80.4 \\
27 & 0.05 & 300.0 & 6.9 & 9.75 & 1 & 8.97 & -51.6 & 225 & -2 & 0.916 & -0.662 & -2.21 & -1.95 & 20.7 & 45.1 & 2.3 & 0.312 & 3.4 & 34.3 \\
28 & 0.05 & 300.0 & 9.2 & 13 & 1 & 8.82 & -30.1 & 242 & -2.1 & 0.998 & -0.624 & -2.44 & -2.07 & 21.6 & 47.6 & 2.76 & 0.374 & 3.74 & 57.9 \\
29 & 0.05 & 400.0 & 6.9 & 9.75 & 1 & 9.44 & -101 & 260 & -3.16 & 1.21 & -1.09 & -3.3 & -3.18 & 24.9 & 50.5 & 1.99 & 0.269 & 4.65 & 22.7 \\
30 & 0.05 & 400.0 & 9.2 & 13 & 1 & 9.33 & -53.8 & 369 & -2.95 & 1.26 & -0.867 & -3.32 & -2.93 & 25.4 & 51.8 & 2.44 & 0.331 & 4.89 & 41.1 \\
31 & 0.0625 & 6.25 & 4.6 & 6.5 & 2 & 8.28 & -6.12 & 19.6 & -0.0385 & 0.216 & 0.104 & -0.059 & 0.261 & 8.25 & 20.6 & 4.77 & 0.751 & 0.601 & 7.36e+03 \\
32 & 0.0625 & 6.25 & 9.2 & 13 & 2 & 6.33 & -19.6 & 20.5 & -0.0877 & 0.332 & 0.151 & -0.123 & 0.359 & 11.4 & 28.5 & 4.9 & 0.772 & 1.45 & 3.67e+04 \\
33 & 0.0625 & 12.5 & 18.4 & 26 & 3 & 3.41 & -114 & 386 & -0.00875 & 0.38 & 0.637 & -0.00268 & 1.01 & 13.4 & 42.1 & 6.56 & 1.03 & 3.74 & 4.51e+05 \\
34 & 0.0625 & 12.5 & 13.8 & 19.5 & 2 & 4.36 & -47.6 & 112 & -0.0546 & 0.374 & 0.341 & -0.0727 & 0.643 & 12.5 & 37 & 5.95 & 0.937 & 2.52 & 1.97e+05 \\
35 & 0.0625 & 200.0 & 9.2 & 13 & 1 & 10 & -21.8 & 212 & -1.76 & 1.04 & -0.447 & -2.17 & -1.57 & 20.2 & 40.8 & 2.9 & 0.457 & 2.87 & 70 \\
36 & 0.0625 & 300.0 & 9.2 & 13 & 1 & 10.3 & -32.5 & 275 & -2.34 & 1.26 & -0.63 & -2.71 & -2.09 & 22.7 & 44.3 & 2.67 & 0.42 & 3.55 & 51.8 \\
37 & 0.0625 & 400.0 & 9.2 & 13 & 1 & 11 & -65.8 & 398 & -3.38 & 1.6 & -0.869 & -3.8 & -3.07 & 26.8 & 48.2 & 2.33 & 0.368 & 4.66 & 35.9 \\
38 & 0.075 & 6.25 & 13.8 & 19.5 & 3 & 5.66 & -82.2 & 101 & -0.0113 & 0.54 & 0.451 & -0.0331 & 0.958 & 15 & 35.3 & 5.65 & 1 & 2.82 & 5.14e+05 \\
39 & 0.075 & 6.25 & 9.2 & 13 & 2 & 10.6 & -39.4 & 24.9 & -0.0419 & 0.691 & 0.286 & -0.0525 & 0.925 & 16.9 & 29 & 4.57 & 0.812 & 1.91 & 3.07e+04 \\
40 & 0.075 & 12.5 & 13.8 & 19.5 & 2 & 6.52 & -116 & 387 & -0.0534 & 0.694 & 0.719 & -0.0714 & 1.34 & 16.7 & 36.4 & 5.21 & 0.927 & 3.04 & 4.75e+05 \\
41 & 0.075 & 12.5 & 18.4 & 26 & 3 & 4.6 & -129 & 109 & -0.00278 & 0.655 & 0.766 & -0.0132 & 1.41 & 17 & 43.5 & 5.68 & 1.01 & 4.46 & 6.34e+05 \\
42 & 0.0875 & 6.25 & 9.2 & 13 & 2 & 12.7 & -36.4 & 24.7 & -0.0661 & 0.926 & 0.282 & -0.0786 & 1.13 & 19.3 & 28.8 & 4.24 & 0.835 & 2.08 & 8.01e+03 \\
43 & 0.0875 & 12.5 & 18.4 & 26 & 3 & 6.54 & -92.7 & 158 & -0.0293 & 1.07 & 1.03 & -0.0529 & 2.06 & 21.6 & 42.3 & 5.89 & 1.16 & 5.08 & 2.13e+05 \\
44 & 0.0875 & 6.25 & 13.8 & 19.5 & 3 & 8.68 & -34.1 & 70.2 & -0.0586 & 0.986 & 0.518 & -0.0903 & 1.41 & 20.5 & 34.8 & 5.62 & 1.11 & 3.44 & 3.68e+05 \\
45 & 0.1 & 6.25 & 13.8 & 19.5 & 3 & 11.3 & -61 & 69 & -0.0608 & 1.46 & 0.711 & -0.122 & 2.05 & 25 & 35.7 & 4.84 & 1.04 & 3.92 & 1.51e+05 \\
46 & 0.1 & 12.5 & 9.2 & 13 & 2 & 13.4 & -72.6 & 107 & -0.119 & 1.09 & 0.415 & -0.134 & 1.37 & 20 & 28.6 & 3.69 & 0.795 & 2.12 & 1.66e+03 \\
47 & 0.1 & 12.5 & 18.4 & 26 & 3 & 9.95 & -95 & -57.2 & -0.97 & 1.88 & 1.22 & -0.988 & 2.11 & 29.2 & 42.4 & 5.06 & 1.09 & 6.11 & 9.02e+04 \\
\enddata

\tablenotetext{a}{{Values of $v_\text{A}/v_\text{esc}$ are a factor of two larger than those quoted in Paper \citetalias{Ireland:2021db}, which printed values evaluated at the equator, rather than the pole.}}
\tablenotetext{b}{{Values of $B_\star$ are taken at $R=R_\star$, which slightly differ from those quoted in Paper \citetalias{Ireland:2021db}, where $B_\star$ was taken at the inner domain (slightly above $R_\star$).}}
\tablenotetext{c}{{Outputted variables for this simulation are time-averages over the first $85 \%$ of the latter half of the simulation's temporal domain, due to an extreme event occurring during the remaining $15 \%$. We leave further exploration of long-time evolution for such extreme systems to future work.}}

\end{deluxetable*}
\begin{deluxetable}{cc}
\tablecaption{Normalization of Each Parameter Type in Table~\ref{tab:Pluto_sims}.\label{tab:conversion}}

\tablehead{Parameter & Normalization}

\startdata
$B_0$ & $(4 \pi \rho_\star v_{\text{K},\star}^2)^{1/2}$ \\
$\dot{M}_0$ & $\rho_\star R_\star^2 v_{\text{K},\star}$ \\
$\dot{J}_0$ & $\rho_\star R_\star^3 v_{\text{K},\star}^2$ \\ 
$\Phi_0$ & $(4 \pi \rho_\star R_\star^4 v_{\text{K},\star}^2)^{1/2}$ \\
\enddata
\end{deluxetable}

\section{Simulations of Star-Disk Interaction}\label{sec:sims}

\subsection{Qualitative Behavior of Star-Disk Interaction}\label{sec:qualitative}

In this section, we focus on the dynamic processes that result from the stellar wind, the accretion flow, and the MEs. We split our analysis into three different regimes: the well-behaved spin-up regime (State 1, which was extensively investigated in Paper \citetalias{Ireland:2021db}), the more-complex, dynamic propeller spin-down regime (State 3), and the transition between these (State 2). We illustrate that these regimes depend on the relative positions of the truncation radius, $R_\text{t}$, the Keplerian corotation radius, $R_\text{co}=f^{-2/3} R_\star$ (the radius at which a Keplerian disk is in corotation with the star)\footnote{Note, $R_\text{co}$ may not be the exact location of where the disk is corotating with the star. Depending on whether the MEs extract or add angular momentum from/to the disk, the true corotation radius could in fact be closer or further from the star, respectively. Since $R_\text{co}$ is known a priori, we adopt this value when categorizing simulations into their respective States, for simplicity.}, and the outermost radial extent of the magnetic connection between the star and the disk, $R_\text{out}$.

\begin{figure*}
\begin{center}
\includegraphics[width=0.49\textwidth]{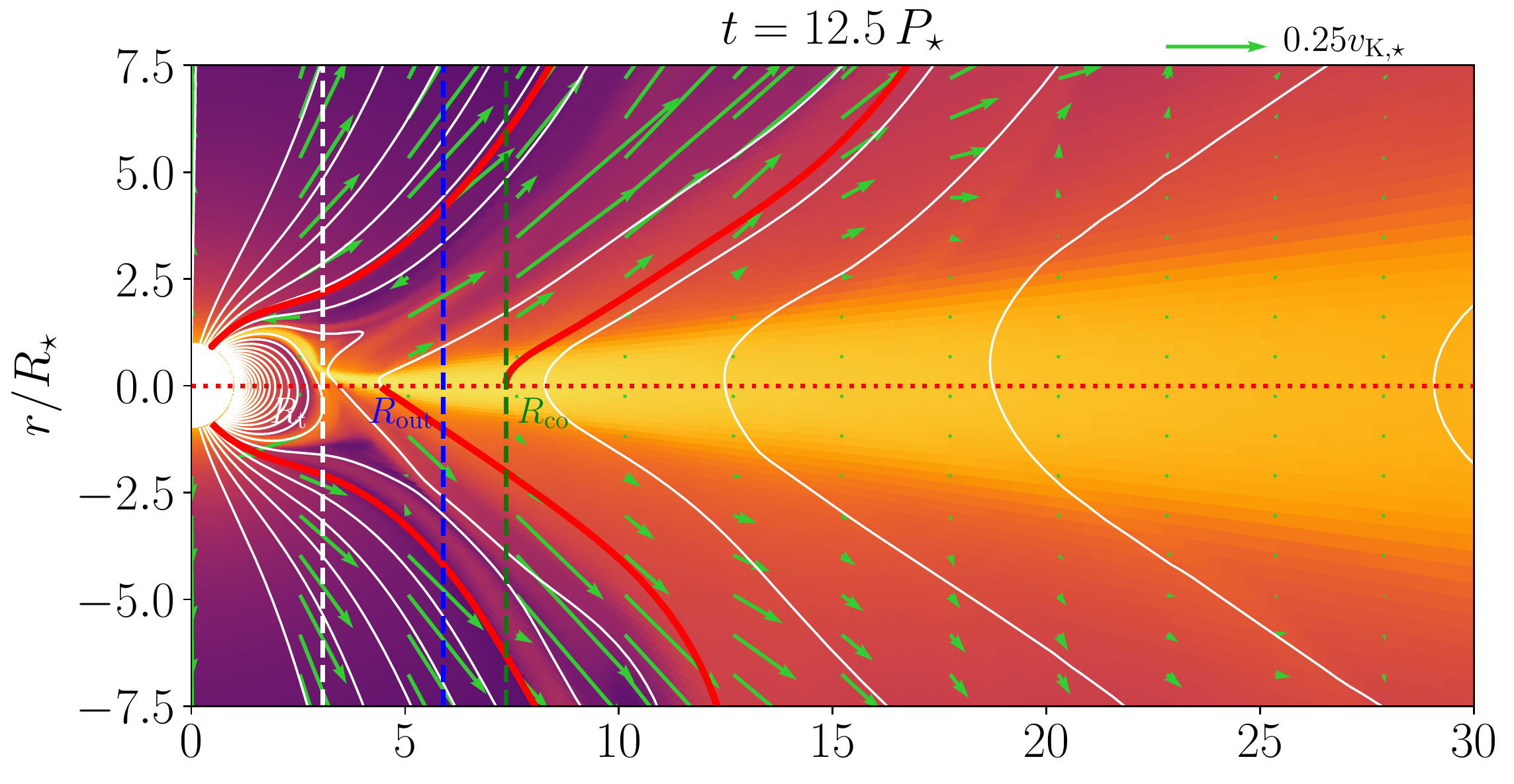}
\includegraphics[width=0.475\textwidth]{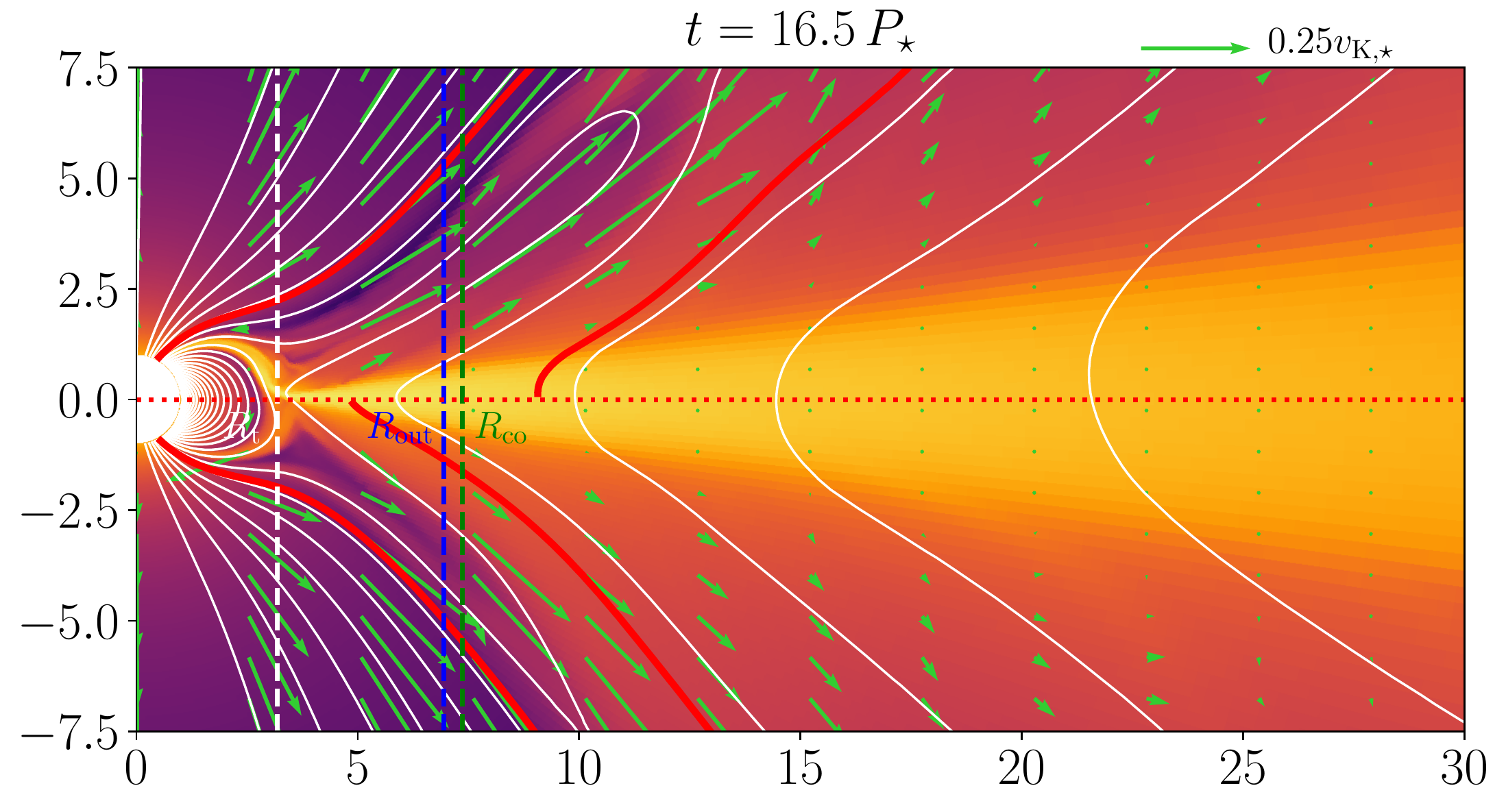}\\
\includegraphics[width=0.49\textwidth]{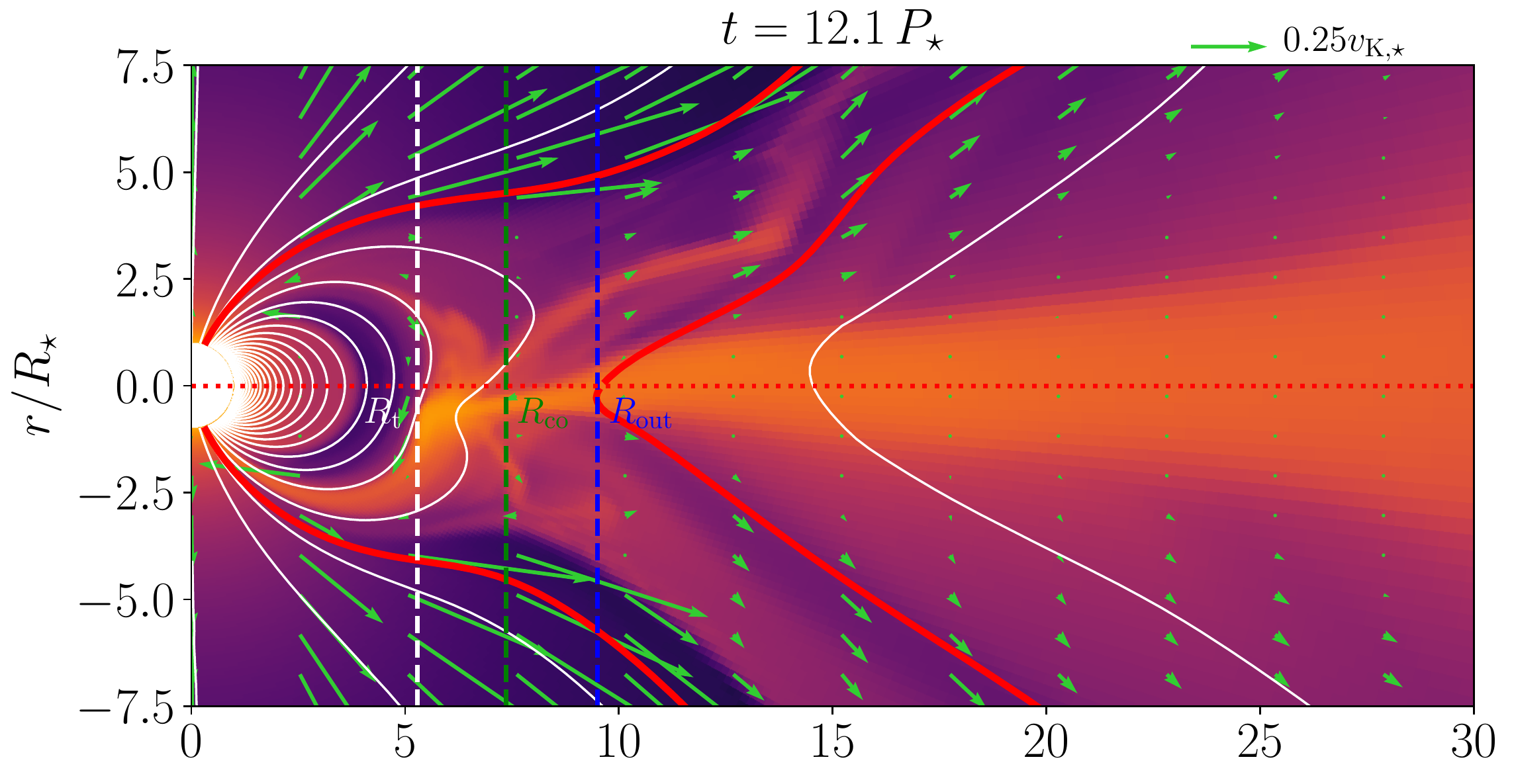}
\includegraphics[width=0.4725\textwidth]{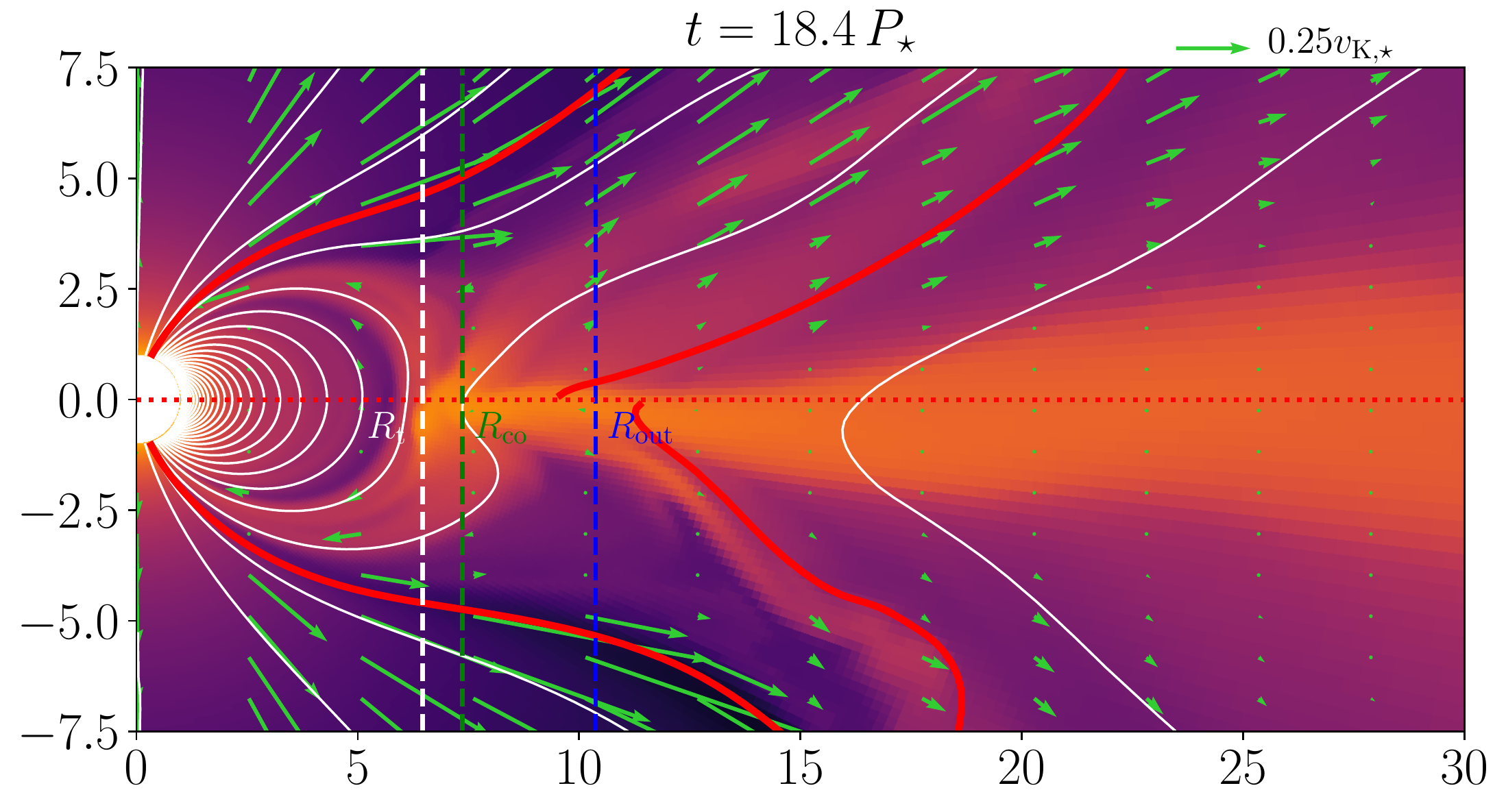}\\
\includegraphics[width=0.49\textwidth]{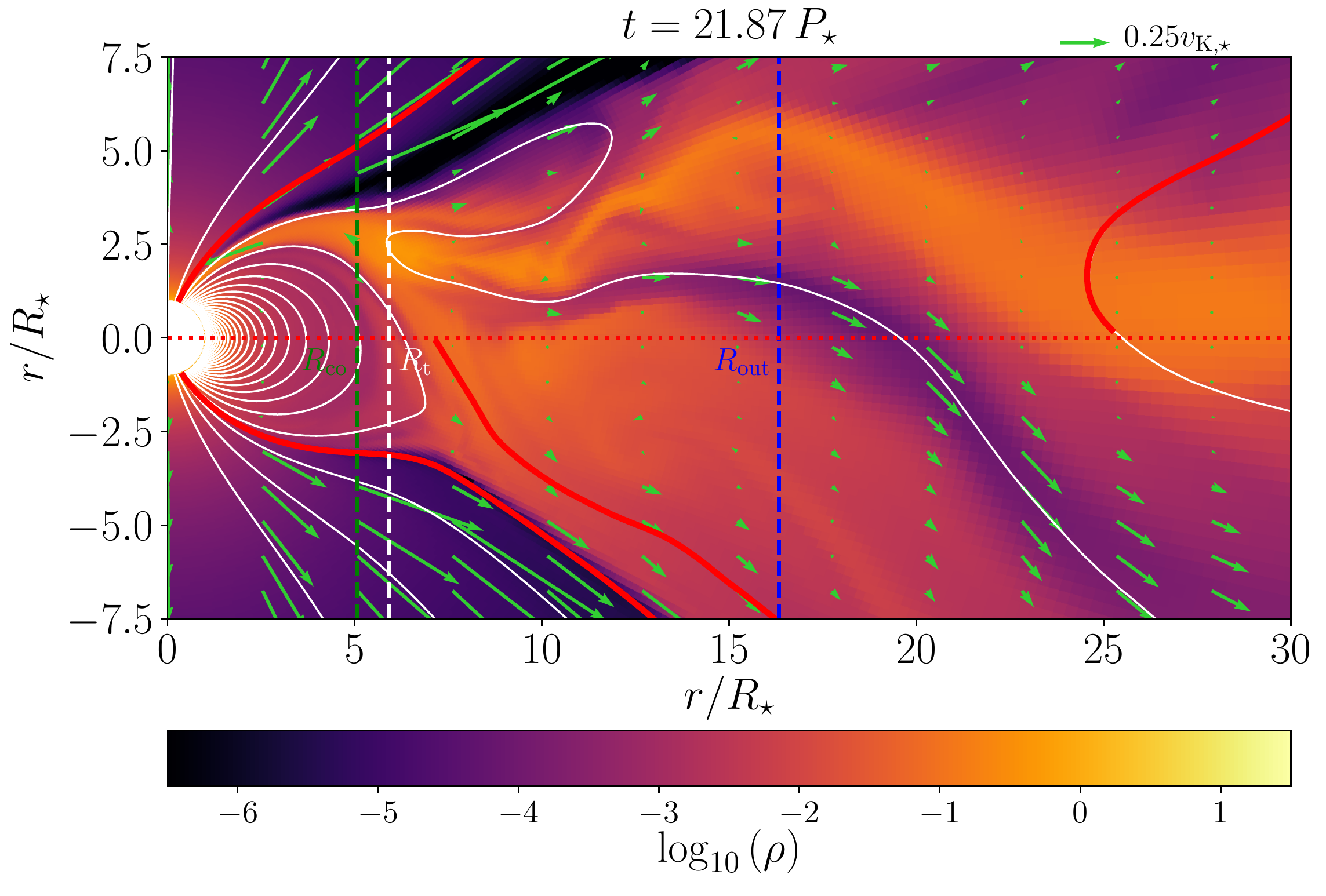}
\includegraphics[width=0.475\textwidth]{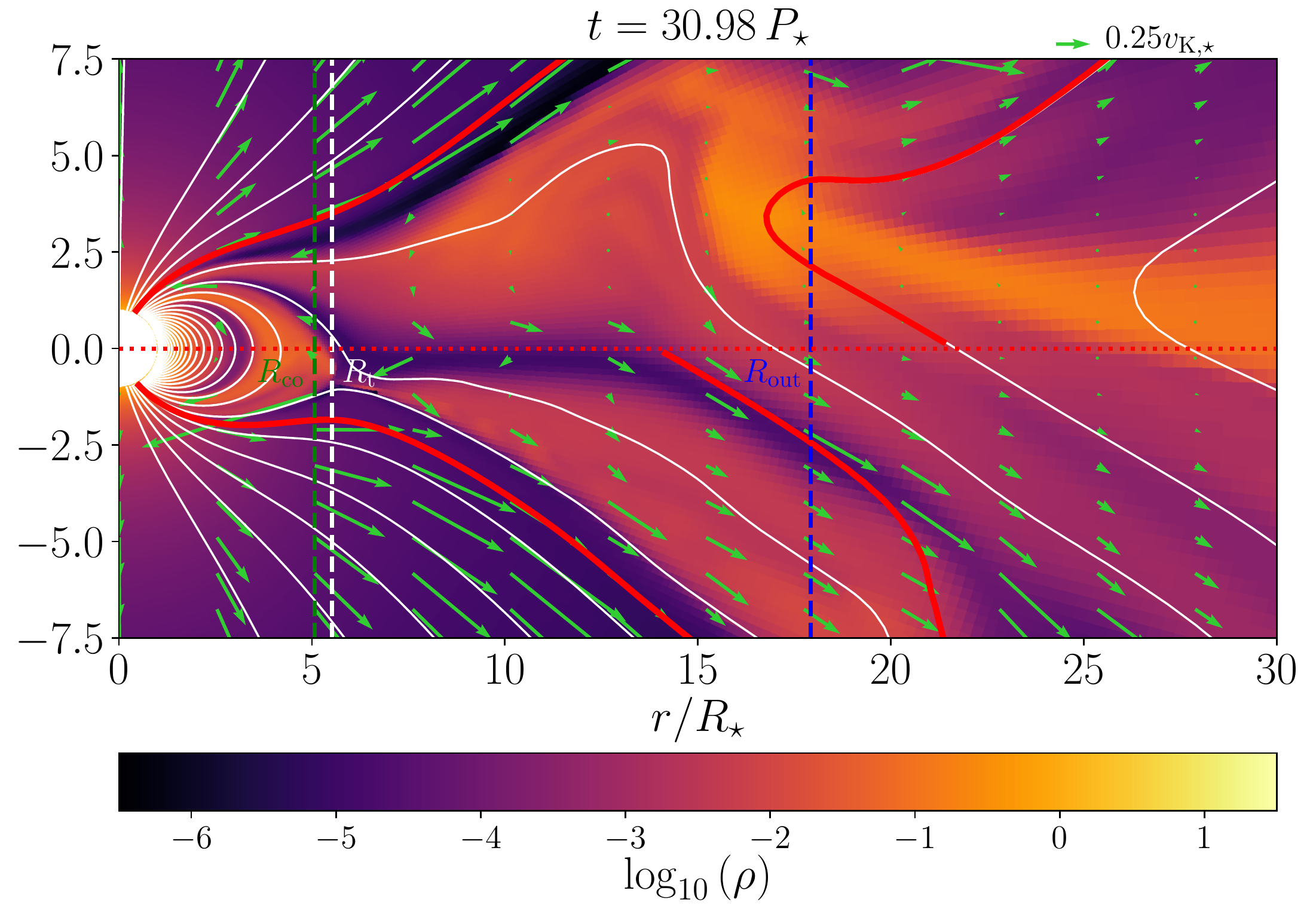}\\
\caption{Logarithmic density color maps, showing the inner domain of (top) State 1 model 26 at time steps (left) 12.5 $P_\star$ and (right) 16.5 $P_\star$, (middle) State 2 model 14 at time steps (left) 12.1 $P_\star$ and (right) 18.4 $P_\star$, and (bottom) State 3 model 43 at time steps (left) 21.87 $P_\star$ and (right) 30.98 $P_\star$. The two time steps shown are to illustrate the level of variation seen in each State. White contours represent magnetic field lines. Red solid lines nearest the poles separate the stellar wind from the MEs. Red solid lines that go through the equator are the same field lines, but here delimit the outer extent of the ME. White dashed line shows the position of $R_\text{t}$, blue dashed line shows the position of $R_\text{out}$, and green dashed line shows the position of $R_\text{co}$. Red dotted line represents the simulation midplane. Green arrows illustrate velocity vectors normalized by $0.25 v_{\text{K},\star}$. \label{fig:density}}
\end{center}
\end{figure*}

In Figure~\ref{fig:density}, we plot density distributions for part of the computational domain of three SDI simulations, representative of States 1, 2, and 3 (top to bottom panels, respectively), taken at two different time steps (left to right panels, respectively). These time steps are chosen to compare contrasting stages of the quasi-periodic behavior in these simulations, which we describe in Sections~\ref{sec:state1}-\ref{sec:state3}. White lines trace magnetic field lines. We dissect the system into three distinct regions: (a) the stellar wind, which is ejected along open field lines anchored to the star, spinning down the star; (b) the MEs, which exhibit the inflation and reconnection of magnetospheric field lines as a result of star-disk differential rotation increasing the toroidal field pressure, either spinning the star up or down (depending on the relative positions of $R_\text{t}$, $R_\text{out}$, and $R_\text{co}$); (c) the accretion flow, where disk material falls from the truncation region onto the stellar surface via accretion funnels, spinning up the star. Region (c) also contains the ``dead zone", where coronal gas corotates around the magnetic equator below the accretion flow.

Our criterion for defining the boundary between the stellar wind and the MEs (red solid lines) is identical to that described in Paper \citetalias{Ireland:2021db}. However, as simulations transition into the propeller regime, it is essential to modify the criterion for defining the truncation radius, $R_\text{t}$ (white dashed line). As in Paper \citetalias{Ireland:2021db}, if disk material is actively accreting onto the star, we define $R_\text{t}$ as the radial coordinate that corresponds to the minimum specific entropy (a proxy for the densest disk material region) along the outermost closed field loop not inflated by the ME phenomenon \citep{Pantolmos2020}. However, if accretion onto the stellar surface is greatly inhibited for a given time step, i.e., the outermost closed field loop instead corresponds to the ``dead zone", we then define $R_\text{t}$ as the radial coordinate corresponding to the equatorial radius of this particular field line. The criterion for defining $R_\text{out}$ (blue dashed line) takes the average equatorial radius between both hemispheres {(as accretion funnel flow can break the midplane symmetry)} of the stellar wind/ME boundary field (red solid) lines that cross the midplane. The green dashed line shows the position of $R_\text{co}$.

\subsubsection{State 1: $R_\text{t}< 0.433 R_\text{co}$}\label{sec:state1}

State 1 applies to all systems where the truncation radius is well within the corotation radius. Explicitly, it is defined by the regime in which the star-disk magnetic connection is within corotation, i.e., $R_\text{t} < R_\text{out} < R_\text{co}$. In this State, the SDI system achieves a quasi-steady configuration; a more detailed description of State 1 can be found in Paper~\citetalias{Ireland:2021db} and \citet{refId0}. In the top panels of Figure~\ref{fig:density}, we plot the density distribution for part of the computational domain of a representative State 1 SDI simulation (model 26), taken at two time steps corresponding to 12.5 and 16.5 stellar rotation periods (left and right panels, respectively). This simulation is initialized with disk density $\rho_{\text{d},\star}/\rho_\star=200$, field strength $B_\star/B_0=13$, and break-up fraction $f=0.05$; this relatively dense accretion disk results in a high dynamical (thermal plus ram) disk pressure experienced by the magnetosphere, pushing the disk truncation inward toward the stellar surface.

For the accretion flow, material is steadily accreting onto the star in each panel, and the location of disk truncation remains roughly constant. The panels demonstrate the quasi-periodic behavior of the MEs, showing an instance of the minimum and maximum extent of the magnetically-connected region between the star and disk (i.e., the distance between $R_\text{t}$ and $R_\text{out}$), respectively. {Our simulations do not assume or enforce north–south symmetry during the evolution. Our current results, as also seen in the simulations of Paper~\citetalias{Ireland:2021db} and \citet{Pantolmos2020}, suggest that an equatorially symmetric solution likely represents an unstable equilibrium. The accretion flow tends to select a hemisphere (either north or south) to form the accretion funnel, with the closed magnetosphere against which it leans slightly displaced on the opposite side. It is also possible for the dominant accretion funnel to switch between hemispheres one or more times during the simulation. In model 26, the asymmetry of the accretion flow (predominantly in the northern hemisphere throughout this particular simulation)} also produces hemispheric asymmetry in the MEs, illustrated by the relative size of this ME region between each panel. In all 22 State 1 simulations, disk material is magnetically-connected to the star within corotation (where the disk is rotating faster than the star), therefore the MEs exert a net spin-up torque contribution onto the star. The extraction of angular momentum from the disk causes the disk to rotate at sub-Keplerian speeds in the vicinity of truncation. The area of the stellar wind remains fairly constant, as the opening angle of the wind (and the fraction of open flux in the wind) is predominantly a function of the truncation radius (as demonstrated in Paper~\citetalias{Ireland:2021db}).

We find $R_\text{out}\approx2.31 R_\text{t}$ (on average) in this regime, therefore (by setting $R_\text{out} = R_\text{co}$) we can approximately define State 1 as the regime where $R_\text{t}< 0.433 R_\text{co}$ is satisfied.

\subsubsection{State 2: $0.433 R_\text{co}<R_\text{t}<R_\text{co}$}

State 2 defines the regime in which $R_\text{t} < R_\text{co} < R_\text{out}$, which we can also write as $0.433 R_\text{co}<R_\text{t}<R_\text{co}$. In this transition State, the SDI system departs from the quasi-steady behavior found in State 1, and becomes increasingly more complex and dynamic as $R_\text{t}$ approaches $R_\text{co}$. The outer extent of the (time-average) magnetically-connected region between the star and disk ($R_\text{out}$) now exceeds the corotation radius, and, as a result, it is possible for the MEs to either spin the star up or down, depending on the relative position of $R_\text{t}$ and $R_\text{out}$, about $R_\text{co}$. In the middle panels of Figure~\ref{fig:density}, we plot the density distribution for part of the computational domain of a representative State 2 SDI simulation (model 14), shown at two times corresponding to 12.1 and 18.4 stellar rotation periods (left and right panels, respectively). This simulation is initialized with an identical field strength $B_\star/B_0=13$ and break-up fraction $f=0.05$ to the State 1 example, but with a much lower disk density $\rho_{\text{d},\star}/\rho_\star=6.25$; in general, this reduction in disk density decreases the dynamical disk pressure experienced by the magnetosphere, pushing the disk truncation outward toward the corotation radius.

Accretion flow is more intermittent due to the truncation of the disk being in closer vicinity to the corotation radius; the two times plotted are chosen to illustrate this cyclic accretion behavior. 
In the first frame, material is temporarily accreting onto the stellar surface, due to an increase in disk density near this boundary. This increases the dynamical disk pressure experienced by the magnetosphere, and pushes the truncation radius inward by a small fraction (further from corotation).
In the second frame, a drop in disk density causes the truncation radius to be pushed outward toward corotation, causing accreting material to build up in the vicinity of the centrifugal barrier, with a fraction being ejected out of the domain. This behavior is quasi-periodic, resulting in cyclic accretion.
In the second frame, the MEs demonstrate greater hemispheric asymmetry, where the opened-closed oscillations of the MEs occur in the northern and southern hemispheres out of phase with each other or at slightly different oscillation frequencies.
For model 14, a majority of the disk material is magnetically-connected to the star outside of corotation (where the disk is rotating slower than the star), and we find that the MEs exert a net spin-down torque contribution onto the star. In addition, the stellar wind appears to be fairly steady between each panel, likely because the change in truncation radius is relatively small.

\subsubsection{State 3: $R_\text{t}> R_\text{co}$}\label{sec:state3}

State 3 defines the regime in which $R_\text{t}> R_\text{co}$. In this State, the SDI system exhibits even more dynamic behavior. Accretion flow is inhibited further, due to the close proximity of $R_\text{t}$ and $R_\text{co}$, causing huge build ups and periodic ejections of material. {In some instances, the disk is noticeably perturbed, with a majority pushed away by the centrifugal barrier, almost disconnecting the star and disk and greatly inhibiting accretion. This is not unique to State 3 (also occurring in State 2), but this behavior is observed to be more frequent and/or more violent in this State.} In this State, the extent of the star-disk magnetic connection completely exceeds corotation (where the disk is rotating slower than the star), and the MEs exert a large spin-down torque contribution onto the star. In the bottom panels of Figure~\ref{fig:density}, we plot the density distribution for part of the computational domain of a representative State 3 SDI simulation (model 43), taken at two time steps corresponding to 21.87 and 30.98 stellar rotation periods (left and right panels, respectively). This simulation is initialized with a stronger field strength $B_\star/B_0=26$, a higher break-up fraction $f=0.0875$, and a slightly higher disk density $\rho_{\text{d},\star}/\rho_\star=12.5$, compared to the State 2 example. In this case, the magnetospheric pressure experienced by the accretion disk is strong enough to push the disk truncation further out; combined with the fact that the star is rotating more rapidly (bringing $R_\text{co}$ closer to the star), the truncation radius is pushed completely outside the corotation radius.

The two time steps plotted again illustrate the cyclic accretion behavior, as well as the strong ME asymmetry. It is evident that the combination of strong magnetospheric pressure (from high field strengths and low disk densities) and rapid rotation rates, required for a State 3 configuration, results in highly dynamic and powerful ME events. For model 43, the size of the stellar wind region increases between each panel, predominantly as a result of the deflation of both the accretion and ME regions. 

It is possible for simulations to switch between States 2 and 3, if $R_\text{t}$ oscillates about $R_\text{co}$. However, for the purposes of this paper, each simulation is categorized according to their time-average values.

\subsection{Global Flow Quantities and Efficiencies}

As in Paper \citetalias{Ireland:2021db}, we investigate the following global flow properties for each flow component (accretion, stellar wind, and MEs): the mass flow rate, torque, and unsigned magnetic flux. We define a surface $S$ for a given radius $R$, and express these quantities as

\begin{equation}\label{eq:mass_rate_int}
\dot{M} = \int_{S} \rho \boldsymbol{v_\text{p}} \cdot \text{d}\boldsymbol{S} = 2\pi R^2 \int_{\theta_1}^{\theta_2} \rho v_R \sin{\theta} \, \text{d}\theta,
\end{equation}

\begin{align}\label{eq:torque_int}
\begin{split}
\dot{J} = {}&
\int_{S} r \left(\rho v_\phi \boldsymbol{v_\text{p}} - \frac{B_\phi \boldsymbol{B_\text{p}}}{4 \pi} \right) \cdot \text{d}\boldsymbol{S} \\
 = {}& 2 \pi R^3 \int_{\theta_1}^{\theta_2} \left(\rho v_\phi v_R - \frac{B_\phi B_R}{4 \pi} \right) \sin^2{\theta} \, \text{d} \theta,
\end{split}
\end{align}
and

\begin{equation}\label{eq:unsigned_flux}
\Phi = \int_{S} \lvert\boldsymbol{B} \cdot \text{d}\boldsymbol{S}\rvert = 2 \pi R^2 \int_{\theta_1}^{\theta_2} \lvert B_R \sin{\theta} \rvert \, \text{d}\theta,
\end{equation}
respectively, where $r=R \sin{\theta}$ is the cylindrical radius, $v_R$, $v_\phi$, and $v_\text{p}$ are the radial, toroidal, and poloidal velocity, respectively, and $B_R$, $B_\phi$, and $B_\text{p}$ are the radial, toroidal, and poloidal magnetic field strength, respectively. 

We investigate these quantities at the stellar surface (determined, in theory, by evaluating Equations~(\ref{eq:mass_rate_int})-(\ref{eq:unsigned_flux}) at $R = R_\star$), and split them into additive contributions from the stellar wind, accretion, and MEs, by integrating these quantities between the two angles, $\theta_1$ and $\theta_2$, that correspond to the angles enclosing the area for each component. In practice, we take the median value of the mass flow rate and torque between $R_\star$ and $2R_\text{t}/3$, rather than at $R_\star$, to avoid numerical effects near the inner boundary. Furthermore, the simulations are not a steady-state, so these formulations calculate an instantaneous flow rate through a given surface; therefore, computing these over a small range of radii produces global quantities analogous to small time-average values. The unsigned stellar magnetic flux is calculated at $R_\star$ as the field strength is fixed at its initial dipolar configuration, and can be expressed analytically as $\Phi_\star = \Phi_\text{wind} + \Phi_\text{acc} + \Phi_{\text{ME},\star} \equiv \alpha \pi R_\star^2 B_\star$ ($\alpha = 2$ for a dipolar surface configuration).
Here, we define the inflow of mass flux (and torque) toward the star as negative quantities, which acts to add mass to the star and spin it up, and vice versa.

\begin{figure*}
\begin{center}
\includegraphics[width=0.333\textwidth]{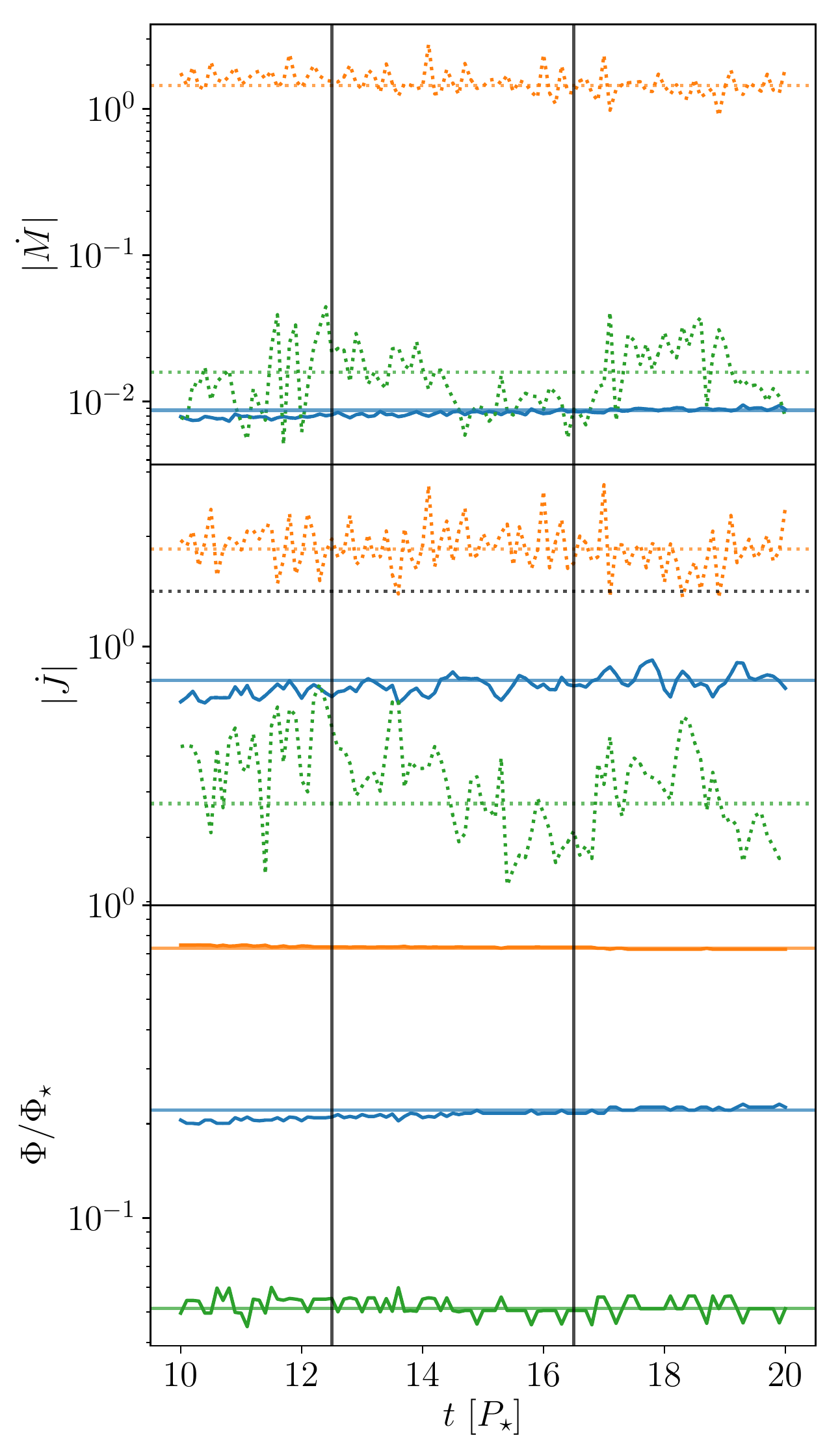}
\includegraphics[width=0.31275\textwidth]{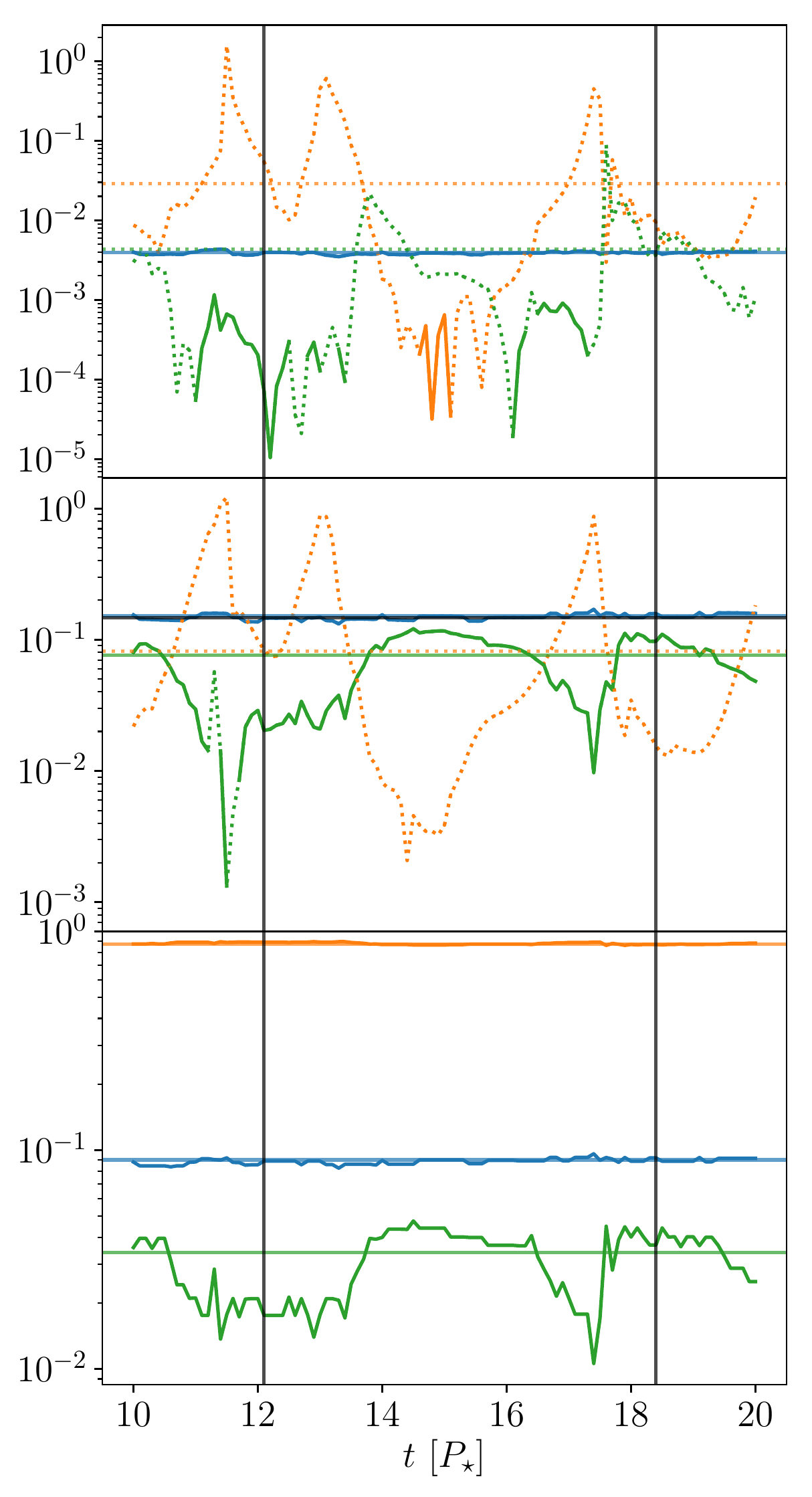}
\includegraphics[width=0.31275\textwidth]{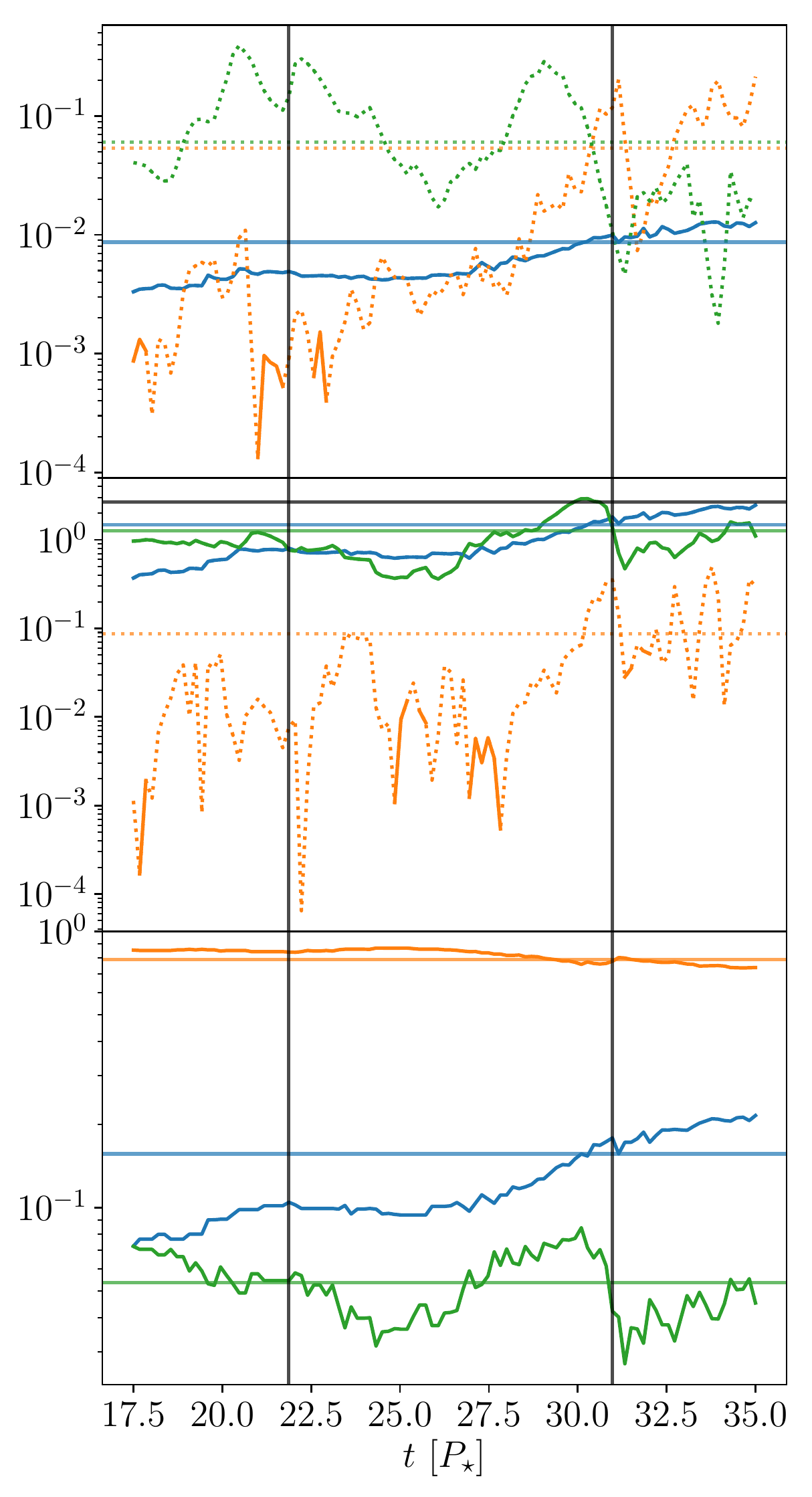}

\caption{Evolution of the stellar mass flow rate, torque, and normalized unsigned magnetic flux (from top to bottom), for the stellar wind (blue), accretion (orange), and MEs (green), as a function of time in units of stellar period, $P_\star$, for (left) State 1 model 26, (center) State 2 model 14, and (right) State 3 model 43. Dotted lines represent negative values, and solid lines represent positive values. Respectively colored horizontal lines indicate time-average values. Black horizontal lines represent the time-average net stellar torque ($\dot{J}_\star$). Black vertical lines indicate the time steps plotted in Figure~\ref{fig:density}. \label{fig:global_evo_v_time}}
\end{center}
\end{figure*}

Figure~\ref{fig:global_evo_v_time} demonstrates the temporal evolution of the separate contributions to the mass flow rate, torque, and unsigned magnetic flux (normalized by $\Phi_\star$), showing models 26, 14, and 43 as examples representative of States 1, 2, and 3, respectively. 
We plot the latter half of the total timescale for each simulation, where simulations have appeared to relax into a quasi-steady or quasi-periodic state, following a transient adjustment from the initial state.
We take time-average values of these global parameters over the plotted timescale (horizontal lines), allowing us to compute flow properties integrated over any periodic (or quasi-periodic) behaviour. Time-averaged global quantities for each simulation can be found in Table~\ref{tab:Pluto_sims}. 

In State 1 (left panels), the accretion (orange) and stellar wind (blue) mass-loss rates, torques, and magnetic fluxes are all quasi-steady, and the MEs (green) demonstrate quasi-periodic behavior in all three quantities over a function of time. A significant proportion of the torque on the star comes from accreting material, thus the star has a strong spin-up torque.
In State 2 (middle panels), both the accretion mass flow rate and torque are highly variable, due to the cyclic accretion behavior. As in State 1, the MEs demonstrate quasi-periodic behavior in all three quantities, and the stellar wind remains quasi-steady. In this simulation, the time-average accretion torque is comparable to the spin-down ME torque.

In general, we expect that the mass flux in the disk (entering from the outermost magnetically-connected region, $R_\text{out}$), $\dot{M}_{\text{disk},R_\text{out}}$, is equal to the sum of the mass flux that falls onto the star via accretion and MEs, the mass flux ejected by the MEs out of the domain, $\dot{M}_{\text{ME,out}}$, and other time-variable behavior of the mass (e.g., build up at the inner disk), $d(M_\text{disk})/dt$. Therefore, for the time-averaged domain, the mass flux balance of a quasi-steady-state configuration ($d(M_\text{disk})/dt \approx 0$) follows

\begin{equation}\label{eq:mass_balance}
\dot{M}_{\text{disk},R_\text{out}} = \dot{M}_{\text{acc}} + \dot{M}_{\text{ME},\star} - \dot{M}_{\text{ME,out}}.
\end{equation}
{Values of $\dot{M}_{\text{ME,out}}$ for each simulation can be found in Table~\ref{tab:Pluto_sims}.} Our State 1 simulations show good correspondence with Equation~(\ref{eq:mass_balance}) in the time-averaged domain, with a mean percentage difference of $\sim 15 \%$. However, we see more deviation as we transition into the propeller regime (States 2/3), where all but one of the cases lie within a factor of 2, likely due to a number of factors: e.g., the asymmetry of the magnetically-connected star-disk region about the midplane (the location of $R_\text{out}$ is taken as the average equatorial radius between both hemispheres), large time-variability (meaning perhaps $d(M_\text{disk})/dt \neq 0$), and the increasingly complex and dynamic behavior that makes it hard to account for the mass flux accurately in general.

It is evident that global properties are slowly increasing with time in State 3 simulations, further demonstrating the quasi-steady configuration of these simulations. In spite of the slow drift of the simulated quantities during the period of our time averages, we show in Section~\ref{sec:torque_formulation} that the global properties obey strict relationships and allow us still to derive robust relationships between global system properties. {For example, it is possible that the State 3 example in Figure~\ref{fig:global_evo_v_time} could eventually drift and transition into a spin-up configuration (State 2), but we expect that it will then follow the appropriate State 2 torque formulation. There are also brief sign changes and extreme dips in the accretion mass flux for these State 2 and 3 simulations, which generally represents temporary periods where the accretion flow is highly disrupted. These episodes are typically short-lived, occurring on the order of a rotation period, with the timescale increasing slightly for increasingly-extreme cases, i.e., as the magnetospheric pressure experienced by the accretion disk increases.} In addition, and especially during these complicated epochs, our algorithm for identifying which field lines should be identified as accretion flow might misidentify them. This should not affect our results, as time averaging likely minimizes the impact these short-lived issues have, and we always account for the total flows to/from the star, so even if some are misidentified at some times, they are always fully accounted for.

\begin{figure}
\begin{center}
\includegraphics[width=0.4625\textwidth]{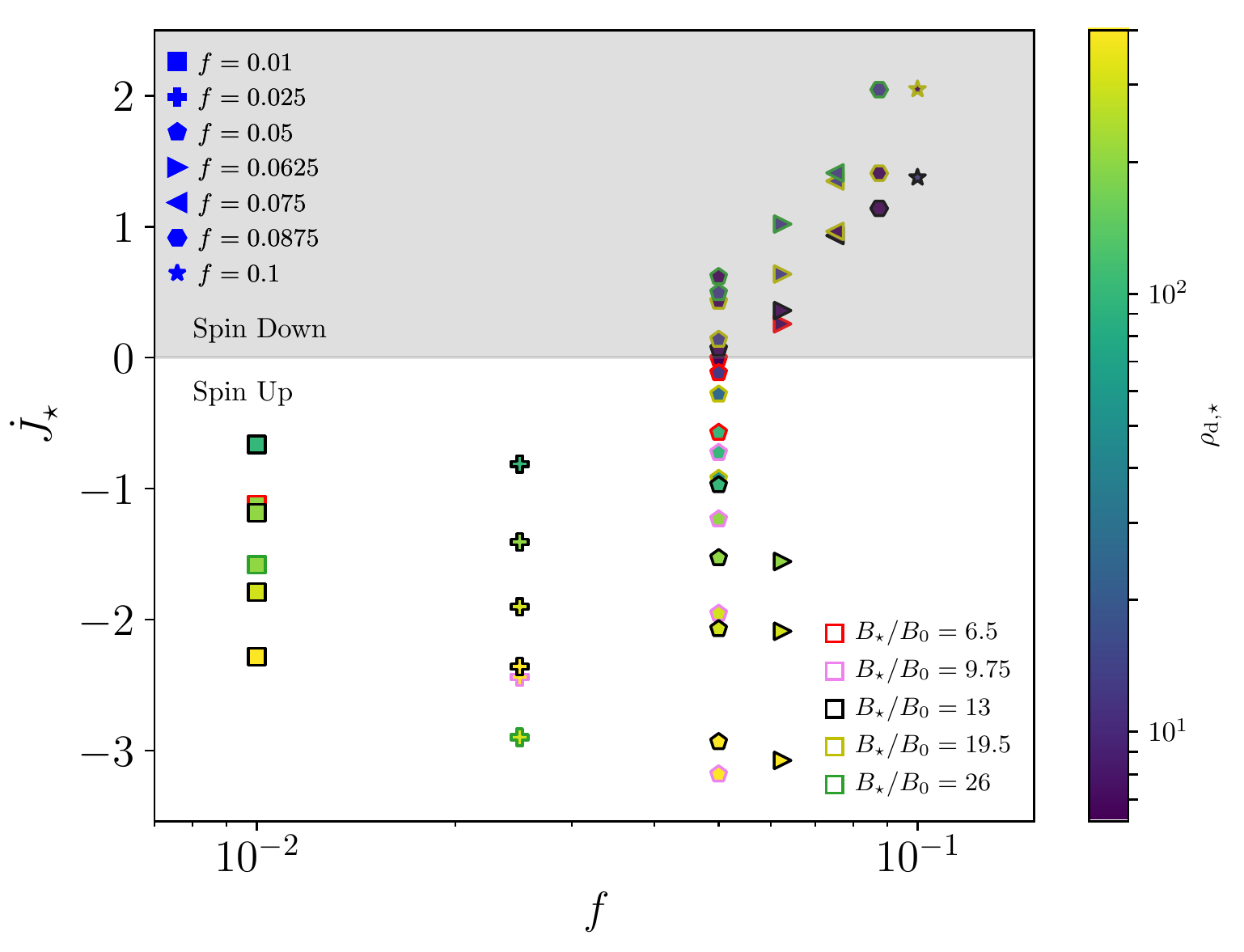}
\caption{The net stellar torque, $J_\star$, as a function of the stellar break-up fraction, $f$, for all simulations. To differentiate between different input parameters, we use differing symbol styles for $f$, marker colors for $\rho_{\text{d,}\star}$, and marker edge colors for $B_\star$. The gray shaded region represents the net spin-down torque regime, whilst the transparent region represents the net spin-up torque regime. \label{fig:Jdot_star_v_f}}
\end{center}
\end{figure}

In Figure~\ref{fig:Jdot_star_v_f}, we plot the net stellar torque, $\dot{J}_\star$, as a function of $f$, to illustrate the extent of the explored parameter space. Symbol shapes, colors, and borders encode the variations of each parameter in this study. In this paper, we push our parameter regime into the spin-down regime, by either increasing $B_\star$, increasing $f$, or decreasing $\rho_{\text{d},\star}$ (or any combination thereof). In Section~\ref{sec:torque_formulation}, we show how these net stellar torque contributions depend on both the stellar and accretion disk global properties in these systems.

\section{Torque Formulations for Simulations of Star-Disk Interaction} \label{sec:torque_formulation}

\subsection{Disk Truncation Radius} \label{sec:Rt_param}

The disk truncation radius, $R_\text{t}$, is defined as the height above the stellar surface where the stresses arising from the stellar magnetosphere disrupts incoming fluid motion from the accretion disk. As with all global parameters in this study, we determine the truncation radius by taking its average value over the latter half of each simulation (unless otherwise stated), where simulations have evolved far beyond the large transient event that occurs as the simulation adjusts from its initial state. These values (normalized by the stellar radius) can be found in Table~\ref{tab:Pluto_sims}. A typical parameterization of the truncation radius for a dipolar magnetic field structure \citep[see, e.g.,][]{2002ApJ...578..420R, 2005MNRAS.356..167M, 2008A&A...478..155B, 2009A&A...508.1117Z} is written as

\begin{equation}\label{eq:Rt_old}
  \frac{R_\text{t}}{R_\star} = K_\text{t} \Upsilon_\text{acc}^{m_\text{t}},
\end{equation}
where 

\begin{equation}\label{eq:Y_acc}
\Upsilon_\text{acc} = \frac{B_\star^2 R_\star^2}{4 \pi \left\lvert \dot{M}_\text{acc}\right\rvert v_\text{esc}},
\end{equation}
is the ``disk magnetization" parameter (ratio of accretion flow's magnetic field and kinetic energies), $\dot{M}_\text{acc}$ is the mass-accretion rate, and $K_\text{t}$ and $m_\text{t}$ are best-fit dimensionless parameters. Values of $\Upsilon_\text{acc}$ for all simulations can be found in Table~\ref{tab:Pluto_sims}. 

\begin{figure}
\begin{center}
\includegraphics[width=0.4625\textwidth]{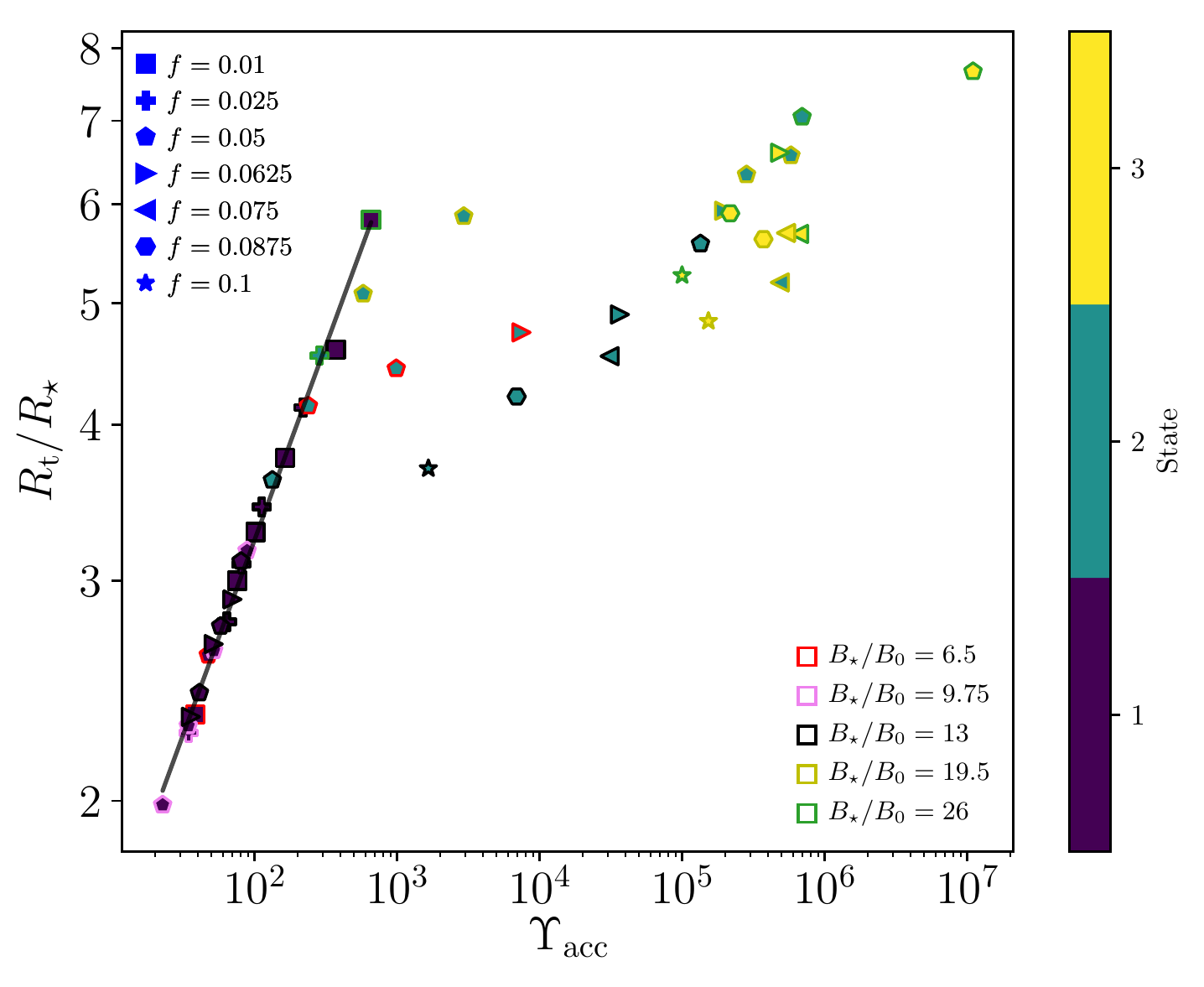}
\caption{Normalized truncation radius, $R_\text{t}/R_\star$, as a function of $\Upsilon_\text{acc}$ for all simulations. Marker colors represent the State of the simulation, hereafter. Black line shows the fit $R_\text{t}/R_\star = 0.772 \Upsilon_\text{acc}^{0.311}$. \label{fig:Rt_v_Upsilon_acc}}
\end{center}
\end{figure}

In Figure~\ref{fig:Rt_v_Upsilon_acc}, we plot the normalized truncation radius as a function of $\Upsilon_\text{acc}$. In State 1, values of $R_\text{t}$ show excellent correspondence with the parameterization expressed in Equation~(\ref{eq:Rt_old}) (as shown in Paper \citetalias{Ireland:2021db}), where $K_\text{t} = 0.772$ and $m_\text{t} = 0.311$. We find a steeper gradient compared to the analytical value $m_\text{t}=2/7$ \citep[e.g.,][]{1977ApJ...215..897E, 2002ApJ...578..420R, 2005MNRAS.356..167M, 2008A&A...478..155B}, as the magnetosphere is perturbed relative to the initial dipolar structure (discussed further in Paper \citetalias{Ireland:2021db}). However, it is clear that this single, power-law relationship no longer correctly predicts $R_\text{t}$ in States~2 and 3 (i.e., as the truncation radius approaches the corotation radius), indicating that the criteria (physical conditions) that determine disk truncation are different in those States.

In States 2 and 3, accretion increasingly becomes inhibited by the centrifugal barrier as $R_\text{t}$ approaches (and eventually exceeds) $R_\text{co}$. As a result, the (time-average) values of mass-accretion rate are significantly less than those present in State 1, where cases are constantly accreting. This can be explained by time-variability, where some of the material that accretes through $R_\text{t}$ will sometimes be subsequently ejected, and sometimes be accreted onto the star. This results in $\Upsilon_\text{acc}$ spanning much greater orders of magnitude for a given $R_\text{t}$, as illustrated in Figure~\ref{fig:Rt_v_Upsilon_acc}. This suggests that the (time-average) truncation radius becomes less sensitive to the magnetization of the disk, and becomes more sensitive to the position of the corotation radius, thus the stellar rotation rate. 

\begin{figure}
\begin{center}
\includegraphics[width=0.4625\textwidth]{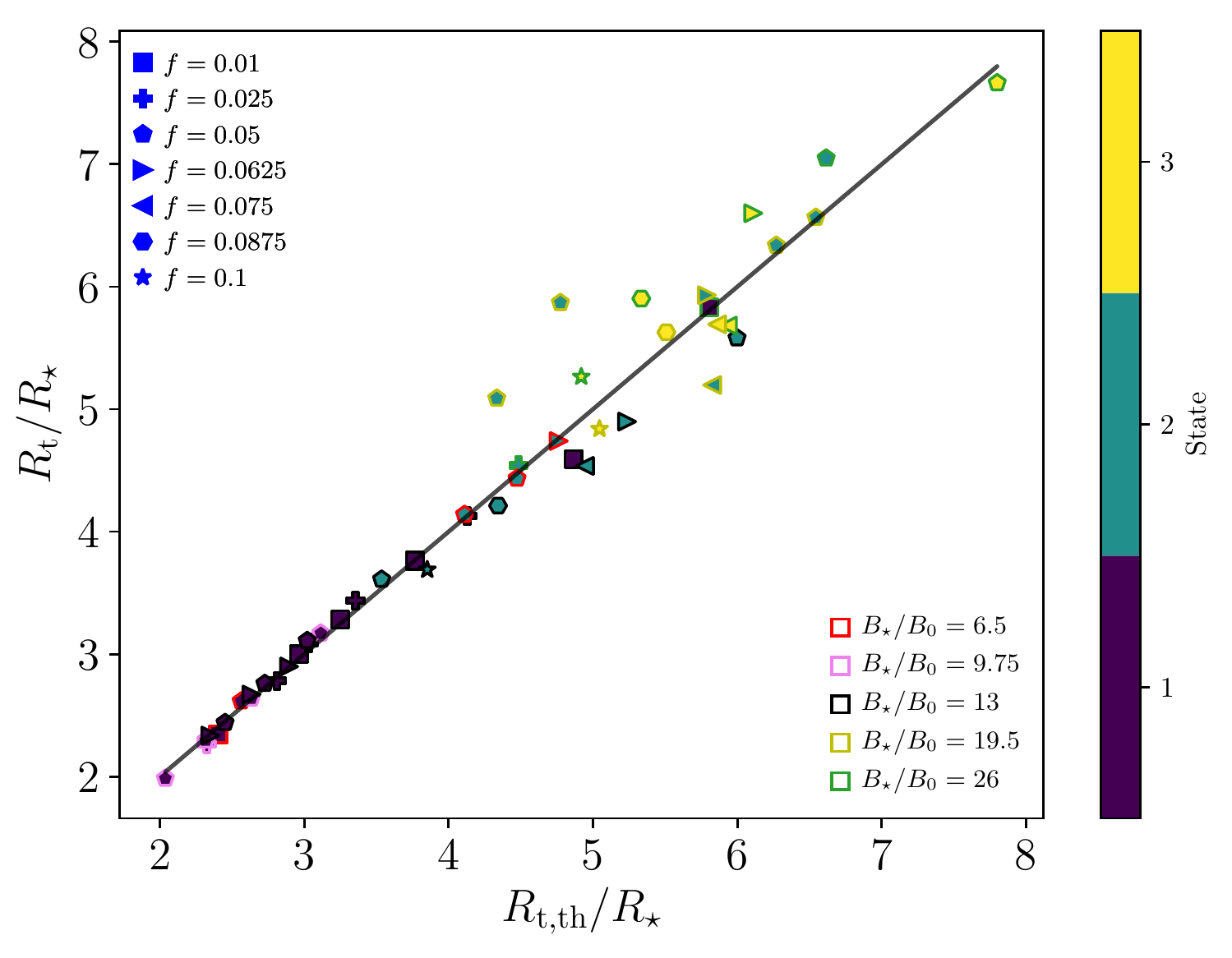}
\caption{Normalized truncation radius, $R_\text{t}/R_\star$, as a function of the parameterization in Equation~(\ref{eq:Rt}) for all simulations. Black line shows $y=x$, indicating goodness of fit. \label{fig:Rt_v_Rt_th}}
\end{center}
\end{figure}

For States 2 and 3, we assume a functional fit for $R_\text{t}$ in terms of $\Upsilon_\text{acc}$ and $f$, for simplicity. We introduce the following parameterization of the normalized truncation radius:

\begin{equation}\label{eq:Rt}
  \frac{R_\text{t}}{R_\star} = \text{min}\{K_{\text{t},1} {\Upsilon_\text{acc}}^{m_\text{t,1}}, K_{\text{t},2} \Upsilon_\text{acc}^{m_{\text{t},2}} f^{m_\text{t,3}}\}
\end{equation}
where $K_{\text{t},i}$ and $m_{\text{t},i}$ are best-fit dimensionless parameters. In Figure~\ref{fig:Rt_v_Rt_th}, we plot $R_\text{t}/R_\star$ as a function of the parameterization expressed in Equation~(\ref{eq:Rt}), where the best-fit dimensionless parameters are $K_{\text{t},1} = 0.772$ and $m_{\text{t},1} = 0.311$ for State 1 simulations, and $K_{\text{t},2} = 1.36$, $m_{\text{t},2} = 0.0597$, and $m_\text{t,3}=-0.261$ for State 2 and 3 simulations. For convenience, all of the best-fit dimensionless parameters presented in Section~\ref{sec:torque_formulation} can be found in Table~\ref{tab:dimensionless_params}. For States 2 and 3, $R_\text{t}$ is much less sensitive to $\Upsilon_\text{acc}$ (compared to State 1) and inversely scales with $f$, i.e., the position of the truncation radius is predominantly influenced by the position of the corotation radius. Due to the simplicity of our functional fit for States 2 and 3, it is likely that we do not accurately capture the complex behavior of $R_\text{t}$ about the transition between States 1 and 2. Thus, we adopt a minimizing function to achieve a continuous transition between these States for our $R_\text{t}$ formulation. We find more scatter with our formulation for States 2 and 3, which is likely due to the simplicity of our functional fit, but also due to the difficulty in determining the truncation radius in these increasingly complex and dynamic systems.

\subsection{Star-Disk Interaction Torque: Accretion and ME} \label{sec:SDI_torque}

In this section, we investigate the torque contribution from the two mechanisms that exert a torque onto the star as a direct result of SDI; namely, the accretion and the MEs. Firstly, the accretion torque accounts for the angular momentum carried by the disk material at $R_\text{t}$, which falls onto the star. It can be given by

\begin{equation}\label{eq:Jdot_acc}
\dot{J}_\text{acc} = \dot{M}_\text{acc} \Omega(R_\text{t}) R_\text{t}^2.
\end{equation}
The analytical value of the accretion torque can be determined by assuming Keplerian rotation, i.e., $\Omega (R_\text{t}) = \Omega_\text{K} (R_\text{t})$, giving $\dot{J}_\text{acc,K} = \dot{M}_\text{acc} (GM_\star R_\text{t})^{1/2}$. However, the accretion disk is not in Keplerian rotation, as the MEs exchange angular momentum with the surface of the disk \citep[see, e.g.,][]{refId0,Pantolmos2020,Ireland:2021db}. Therefore, we rewrite Equation~(\ref{eq:Jdot_acc}) as

\begin{equation}\label{eq:Jdot_acc2}
\dot{J}_\text{acc} = K_\text{acc} \dot{M}_\text{acc} (GM_\star R_\text{t})^{1/2},
\end{equation}
where $K_\text{acc}=\Omega(R_\text{t})/\Omega_\text{K}(R_\text{t})$ is a measure of the ``Keplerianity" of the disk at $R_\text{t}$ ($K_\text{acc} < 1$ for sub-Keplerian, $K_\text{acc} = 1$ for Keplerian, and $K_\text{acc} > 1$ for super-Keplerian).

The impact of the ME torque on the angular momentum of the star-disk system is dynamically more complex, as they exploit magnetic field lines that are connected both to the disk and the star, which are subject to episodes of inflation and reconnection. The MEs can therefore exchange angular momentum both with the disk and the star. From the point of view of the disk, MEs tend to extract angular momentum from it, thus promoting accretion
and favoring sub-Keplerian rotation in the truncation region of the disk, corresponding to a $K_\text{acc} < 1$ factor in Equation~(\ref{eq:Jdot_acc2}). The magnetic torque exerted by MEs onto the disk in the region connected to the star can be customarily written as

\begin{equation}\label{eq:diff_mag_torque}
\dot{J}_{\text{ME,d}} = \int_{R_\text{t}}^{R_\text{out}} q B_\text{p}^2 R^2 \, dR
\end{equation}
where $q=B_\phi^+/B_\text{p}$ is a measure of the field twist, i.e., the ratio of the toroidal and poloidal magnetic field strength at the disk surface. From the point of view of the star, it is usually
assumed that the exchange of angular momentum and therefore the stellar torque due to MEs, $\dot{J}_{\text{ME},\star}$, is mainly determined by a differential rotation effect between the star and the MEs,
whose rotation should reflect the toroidal speed of the connected disk region from which the
MEs are launched. We can therefore expect the stellar ME torque to be made up by a spin-up contribution, $\dot{J}_{\text{ME},\star,\uparrow}$, exerted on the stellar surface magnetically connected to the
disk in the region below $R_\text{co}$ in States 1 and 2, and by a spin-down contribution $\dot{J}_{\text{ME},\star,\downarrow}$,
exerted on the stellar surface connected to the disk in the region beyond $R_\text{co}$ in States 2 and
3. Due to the complex and dynamic nature of the MEs, a formulation for the ME torque on
the star is guided by the physics, but is also heuristic in nature.

Qualitatively, we expect that a fraction of the disk angular momentum extracted from
the connected disk region inside $R_\text{co}$ in States 1 and 2 ($R_\text{t} < R_\text{co}$) can be transferred to the
star, determining a spin-up torque $\dot{J}_{\text{ME},\star,\uparrow}$, the rest being ejected at large distances.
Therefore, we could assume $\dot{J}_{\text{ME},\star,\uparrow}$ to be proportional to the integral
in Equation~(\ref{eq:diff_mag_torque}), taken between $R_\text{t}$ and $\text{min} \{R_\text{out},R_\text{co}\}$. On the other hand, we have found the spin-up
torque exerted in this case to be proportional, within good approximation, to the analytical
accretion torque, so that

\begin{equation}\label{eq:Jdot_ME_star_up}
\dot{J}_{\text{ME},\star,\uparrow} \propto \dot{M}_\text{acc} (GM_\star R_\text{t})^{1/2}.
\end{equation}

Analogously, we can expect the spin-down torque $\dot{J}_{\text{ME},\star,\downarrow}$ exerted by the magnetic surfaces connected to the disk beyond $R_\text{co}$ in States 2 and 3 to enhance the angular momentum flux extracted by the MEs from the
disk. In other words, we assume $\dot{J}_{\text{ME},\star,\downarrow}$ to contribute and be
proportional to the integral
in Equation~(\ref{eq:diff_mag_torque}), taken between $\text{max} \{R_\text{t},R_\text{co}\}$ and $R_\text{out}$. Assuming an average field twist corresponding to some critical value $q = q_\text{c} \sim 1$,\footnote{The SDI system goes through states where the field is maximally twisted and thus open, followed by reconnection. As star-disk differential rotation twists the field rapidly, the twist will actually have an average value close to the maximum value ($q_\text{max}=1$).} and a dipolar magnetosphere $B_\text{p} \propto B_\star (R/R_\star)^{-3}$, we obtain

\begin{align}\label{eq:Jdot_ME_star_down}
\begin{split}
\dot{J}_{\text{ME},\star,\downarrow} {}& \propto \frac{B_\star^2 R_\star^6}{R_i^3} \left[1 - \left(\frac{R_i}{R_\text{out}}\right)^3\right]
\end{split}
\end{align}
where $R_i = \text{max} \{R_\text{co},R_\text{t}\}$.

We can therefore define
the total torque exerted by MEs on the star as

\begin{align}\label{eq:Jdot_ME_star_up_plus_down}
\begin{split}
\dot{J}_{\text{ME},\star} = \dot{J}_{\text{ME},\star,\uparrow} + \dot{J}_{\text{ME},\star,\downarrow},
\end{split}
\end{align}
with the spin-up contribution being present in States 1 and 2 only and the spin-down term
contributing in States 2 and 3 only. Clearly, in States 1 and 3, MEs can only produce
a spin-up or a spin-down torque respectively, whereas in State 2, the sign of the ME torque depends
on the relative position of the truncation and corotation radii. As a last approximation, we
will assume that in State 3 (the ``propeller" regime), most of the disk material is ejected out by
the centrifugal barrier, which also strongly inhibits stellar accretion, i.e., $\dot{J}_\text{acc} \ll \dot{J}_{\text{ME},\star}$. 

We express the total torque exerted on the star by SDI (accretion plus MEs), $\dot{J}_\text{SDI} = \dot{J}_\text{acc} + \dot{J}_{\text{ME},\star}$, as

\begin{equation}\label{eq:SDI_torque}
  \dot{J}_{\text{SDI}} = \begin{cases}
     K_{\text{SDI},1} \dot{M}_\text{acc} (GM_\star R_\text{t})^{1/2} & \text{if $R_\text{t} < 0.433 R_\text{co}$}, \\
K_{\text{SDI},1} \dot{M}_\text{acc} (GM_\star R_\text{t})^{1/2} \\
\hfill + K_{\text{SDI,2}} B_\star^2 R_\star^6 / R_\text{co}^3 & \text{if $0.433 R_\text{co} < R_\text{t} < R_\text{co}$}, \\
     K_{\text{SDI},2} B_\star^2 R_\star^6 / R_\text{t}^3 & \text{if $R_\text{t} > R_\text{co}$},
  \end{cases}
\end{equation}
where $K_{\text{SDI},1}$ and $K_{\text{SDI},2}$ are best-fit dimensionless parameters. For State 2 simulations where $R_\text{out}$ noticeably exceeds $R_\text{co}$, and for all State 3 simulations, we find that $(R_\text{i}/R_\text{out})^3 \ll 1$, removing the $R_\text{out}$ dependency in Equation~(\ref{eq:SDI_torque}). The transition between States 1 and 2 remains approximately continuous regardless of the apparent discontinuity in Equation~(\ref{eq:SDI_torque}), as State 2 simulations near this transition (i.e., where $R_\text{out}$ is close to $R_\text{co}$) have a negligible spin-down ME torque, thus the second additive term can be ignored for these cases. Similarly, the transition between States 2 and 3 happens when the accretion rate is generally low (due to centrifugal barrier), so the first term becomes very small compared to the second term.

\begin{figure}
\begin{center}
\includegraphics[width=0.4625\textwidth]{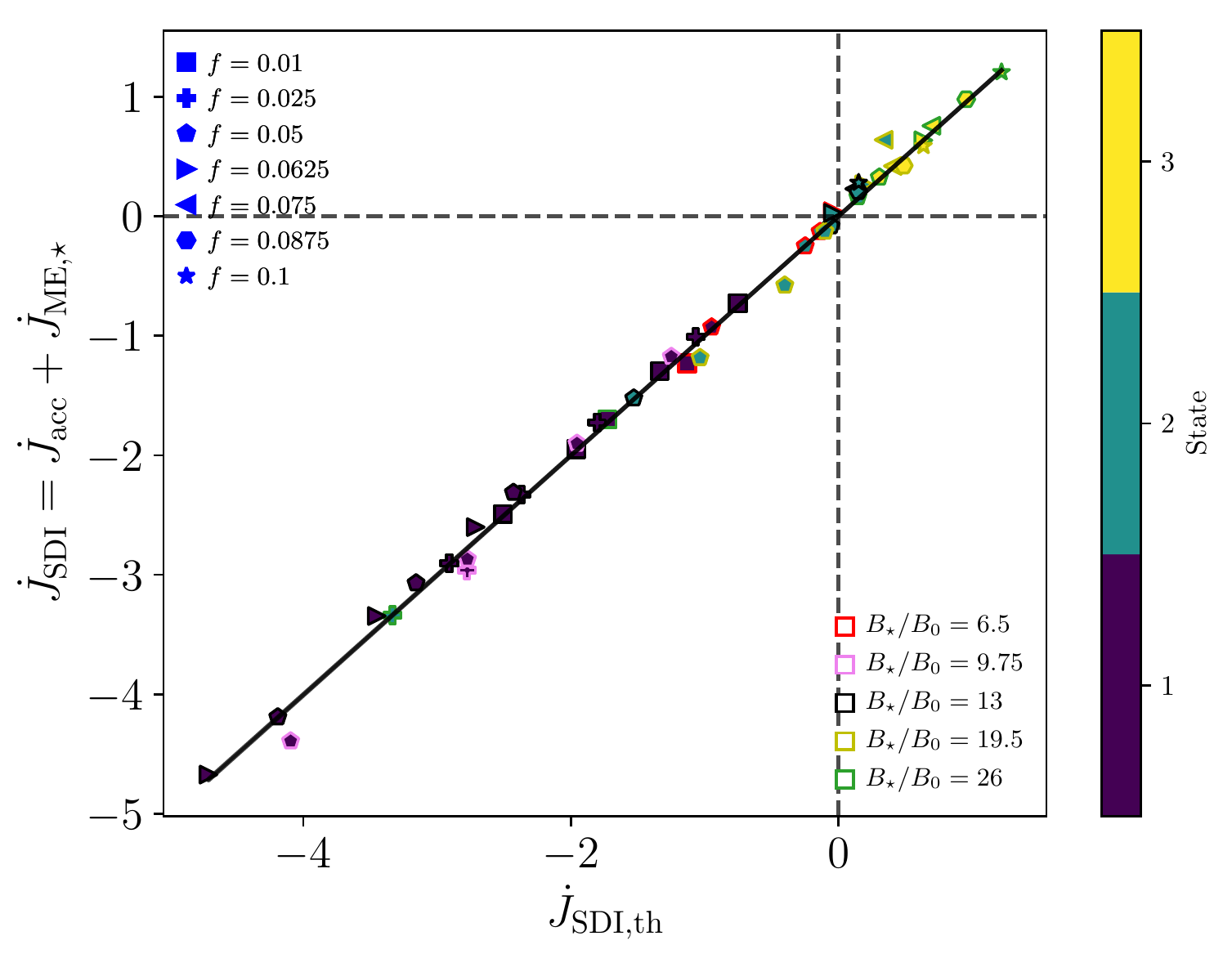}
\caption{SDI torque, $\dot{J}_\text{SDI} = \dot{J}_\text{acc} + \dot{J}_{\text{ME},\star}$, as a function of the parameterization in Equation~(\ref{eq:SDI_torque}) for all simulations. Black line shows $y=x$, indicating goodness of fit. \label{fig:Jdot_SDI_v_Jdot_SDI_theory}}
\end{center}
\end{figure}

In Figure~\ref{fig:Jdot_SDI_v_Jdot_SDI_theory}, we plot $\dot{J}_\text{SDI}$ as a function of the parameterization expressed in Equation~(\ref{eq:SDI_torque}), where the best-fit dimensionless parameters are $K_{\text{SDI},1}=0.909$ and $K_{\text{SDI},2}=0.0171$. In this figure, we use values of $R_\text{t}$ determined by the parameterization expressed in Equation~(\ref{eq:Rt}), instead of the actual simulation values. Changes in the formulation between parameterized and simulated values of $R_\text{t}$ are negligible on the whole, apart from some small deviations in the State 1-2 transition region. However, we choose to adopt the parameterized $R_\text{t}$ values to illustrate that our torque formulation can closely predict the SDI torque a priori.

\subsection{Stellar Wind Torque} \label{sec:SW_torque}

The stellar wind torque is typically parameterized as
\begin{equation}\label{eq:torque_wind}
\dot{J}_\text{wind} = \dot{M}_\text{wind} \Omega_\star \langle r_\text{A} \rangle^2,
\end{equation}
where $\langle r_\text{A} \rangle$ is the torque-average Alfvén radius. The normalized Alfvén radius in our simulations is computed by rearranging Equation~(\ref{eq:torque_wind}):
\begin{equation}\label{eq:RA_torque_wind}
\frac{\langle r_\text{A} \rangle}{R_\star} = \left(\frac{\dot{J}_\text{wind} }{ \dot{M}_\text{wind} \Omega_\star R_\star^2}\right)^{1/2}
\end{equation}
\citep[following][]{1993MNRAS.262..936W,2008ApJ...678.1109M}. This ``effective magnetic lever arm'' quantifies the wind spin-down torque efficiency \citep[see, e.g.,][]{1967ApJ...148..217W}.
Values for $\langle r_\text{A} \rangle / R_\star$ in our simulations can be found in Table~\ref{tab:Pluto_sims}.

\begin{figure}
\begin{center}
\includegraphics[width=0.4625\textwidth]{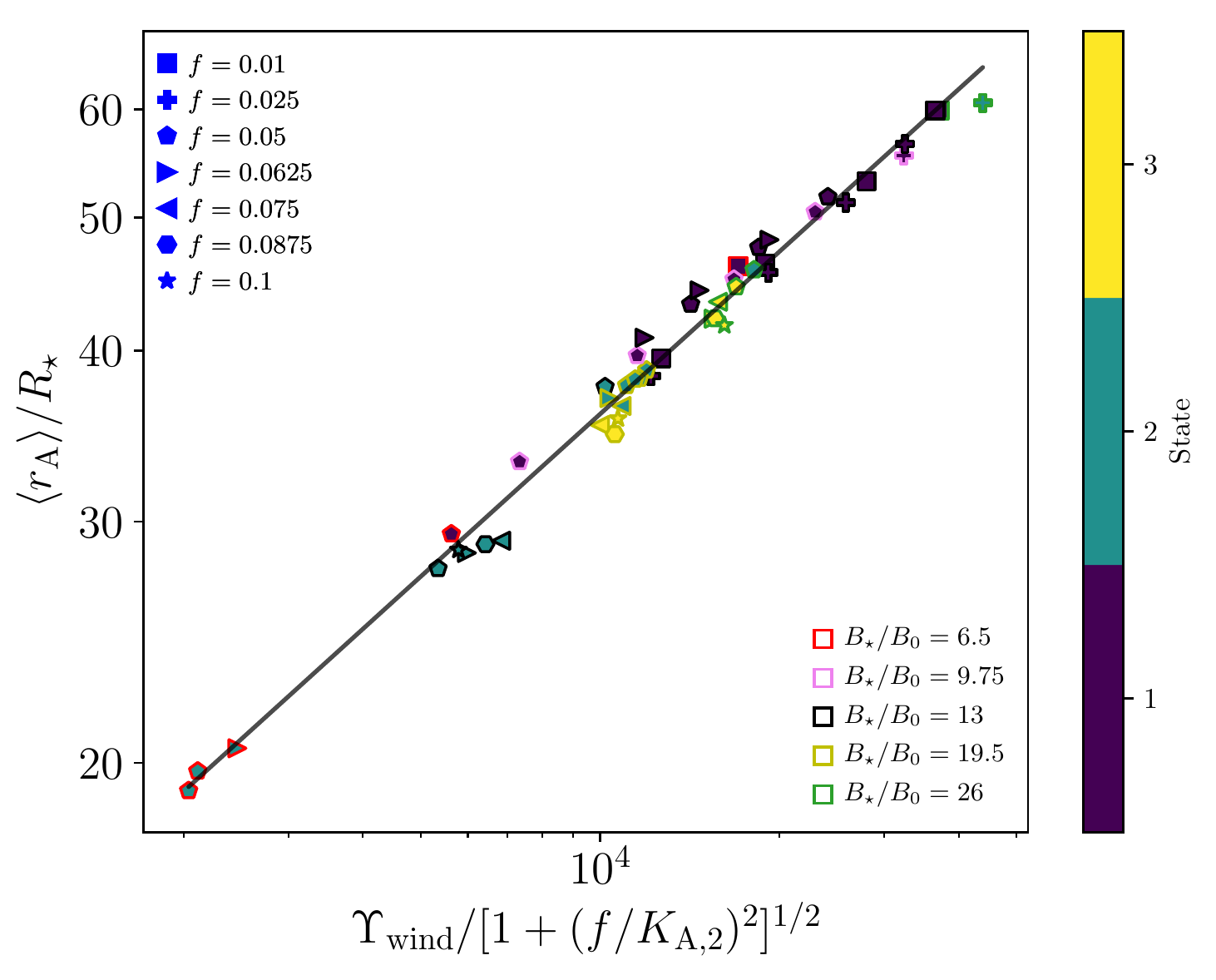}
\caption{Normalized Alfvén radius in the stellar wind, $\langle r_\text{A} \rangle/R_\star$, as a function of the wind flux magnetization parameter, $\Upsilon_\text{wind} / [1 + (f/K_{\text{A},2})^2]^{1/2}$, for all simulations.  The black line shows the fit $\langle r_\text{A} \rangle/R_\star = 0.954 \{\Upsilon_\text{wind} / [1 + (f/0.0284)^2]^{1/2}\}^{0.394}$. \label{fig:rA_v_Upsilon_wind}}
\end{center}
\end{figure}

We use the torque formulation for stellar wind simulations, developed by \citet{2015ApJ...798..116R}, which relates the Alfvén radius to the open magnetic flux in the stellar wind, $\Phi_\text{wind}$, via a single power law that is independent of the magnetic topology:

\begin{equation}\label{eq:Ra_Y_wind}
\frac{\langle r_\text{A}\rangle}{R_\star} = K_{\text{A},1} \left\{\frac{\Upsilon_\text{wind}}{[1 + (f/K_{\text{A},2})^2]^{1/2}}\right\}^{m_\text{A}},
\end{equation}
where $K_{\text{A},1}$, $K_{\text{A},2}$, and $m_\text{A}$ are best-fit dimensionless parameters, and
\begin{equation}\label{eq:Y_wind}
\Upsilon_\text{wind} = \frac{\Phi_\text{wind}^2}{4 \pi R_\star^2\dot{M}_\text{wind} v_\text{esc}}
\end{equation}
is the ``magnetization parameter'' of the stellar wind based on the wind flux. Values for $\Upsilon_\text{wind}$ in our simulations can be found in Table~\ref{tab:Pluto_sims}. Figure~\ref{fig:rA_v_Upsilon_wind} shows $\langle r_\text{A} \rangle/R_\star$ as a function of $\Upsilon_\text{wind} / [1 + (f/K_{\text{A},2})^2]^{1/2}$, where the best-fit dimensionless parameters $K_{\text{A},1}=0.954$, $K_{\text{A},2}=0.0284$, and $m_\text{A}=0.394$. Increasing the wind flux increases the Alfvén radius, giving a more efficient stellar wind torque on the star (i.e., increased torque for a given $\dot M_\text{wind} \Omega_\star$). 
Increasing the stellar wind mass-loss rate, which itself is dependent on the coronal temperature or density, and $\Phi_\text{wind}$, decreases the Alfvén radius.

To predict the stellar wind torque as a function of stellar surface properties, we must parameterize the fraction of flux in the stellar wind. In SDI simulations, star-disk differential rotation results in open magnetic field lines connecting to the star at higher latitudes, compared to where the field connects to $R_\text{t}$ \citep[see, e.g.,][]{2002ApJ...565.1205U}. In Paper \citetalias{Ireland:2021db}, we introduced the following formulation:

\begin{align}\label{eq:phi_star_param}
\begin{split}
\frac{\Phi_\text{wind}}{\Phi_\star} {}& = K_{\Phi,1} \left(\frac{R_\text{t}}{R_\star}\right)^{m_{\Phi,1}} f^{m_{\Phi,2}} \\
& \equiv \left(\frac{\Upsilon_\text{wind}}{\Upsilon_\star}\right)^{1/2},
\end{split}
\end{align}
where $K_{\Phi,1}$, $m_{\Phi,1}$, and $m_{\Phi,2}$ are best-fit dimensionless parameters, and

\begin{equation}\label{eq:Y_star}
\Upsilon_\star = \frac{\Phi_\star^2}{4 \pi R_\star^2\dot{M}_\text{wind} v_\text{esc}}
\end{equation}
is the ``magnetization parameter'' of the stellar wind based on properties at the stellar surface, and $\Phi_\star = \alpha \pi R_\star^2 B_\star$ is the total unsigned stellar magnetic flux ($\alpha=2$ for our dipolar surface configuration). 

\begin{figure}
\begin{center}
\includegraphics[width=0.4625\textwidth]{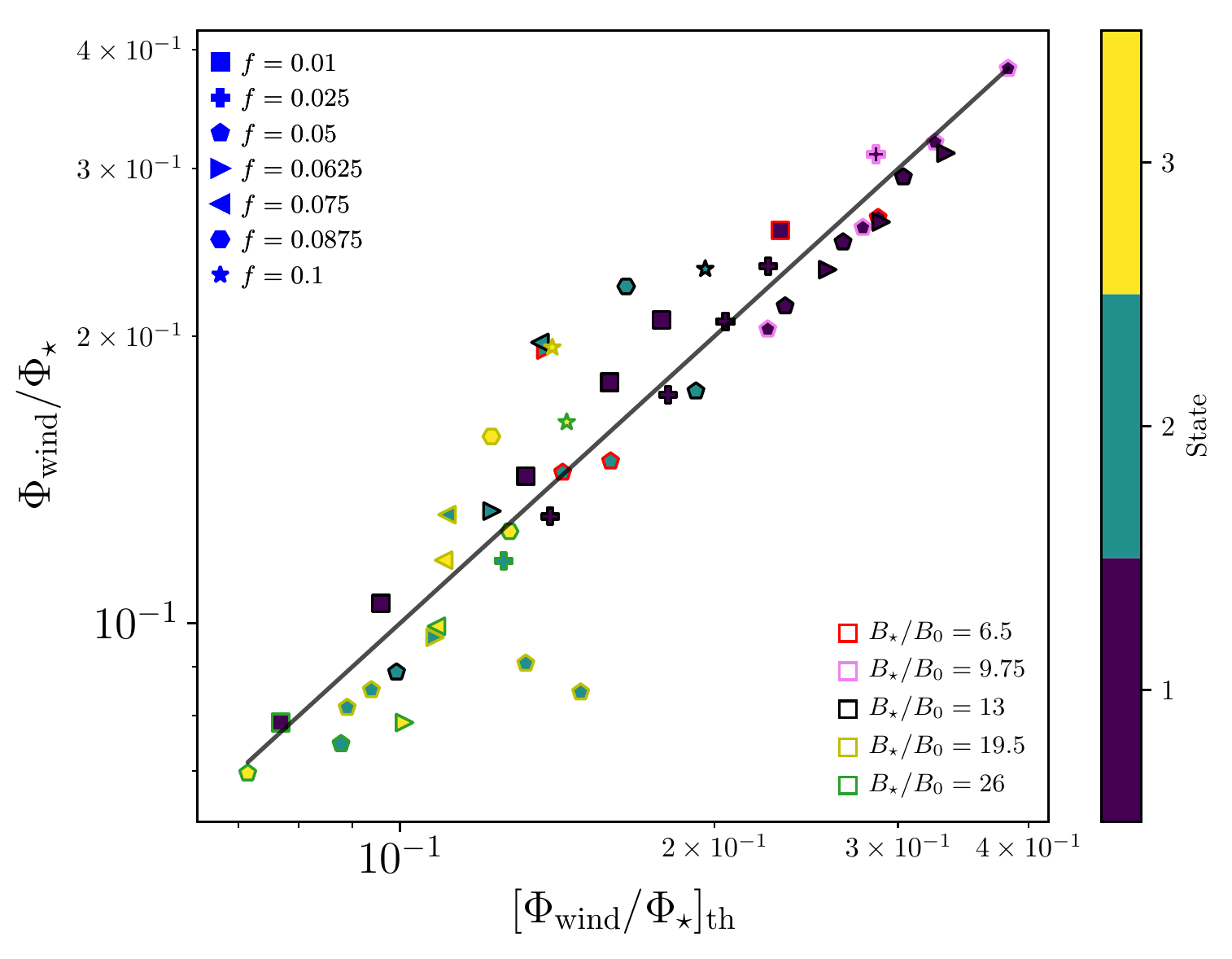}
\caption{Fraction of flux in the stellar wind, $\Phi_\text{wind}/\Phi_\star$, as a function of the parameterization in Equation~(\ref{eq:phi_star_param}) for all simulations. Black line shows $y=x$, indicating goodness of fit. \label{fig:phi_wind_div_star}}
\end{center}
\end{figure}

In Figure \ref{fig:phi_wind_div_star}, we plot $\Phi_\text{wind}/\Phi_\star$ as a function of the parameterization expressed in Equation~(\ref{eq:phi_star_param}), where the best-fit dimensionless parameters are $K_{\Phi,1}=1.62$, $m_{\Phi,1}=-1.25$, and $m_{\Phi,2}=0.184$. Qualitatively, at fixed $f$, a smaller truncation radius results in field lines connecting to the star at lower latitudes, which increases the wind flux by expanding the area on the stellar surface for stellar wind ejection. For a fixed $R_\text{t}$, more rapid stellar rotation decreases the star-disk differential rotation, reducing the fraction of flux in the MEs, which opens up more flux in the stellar wind. 

In this figure, we again use values of $R_\text{t}$ determined by the parameterization expressed in Equation~(\ref{eq:Rt}), instead of the actual simulation values. For State 1, we see good correspondence between simulation values of $\Phi_\text{wind}/\Phi_\star$ and our parameterization. For States 2 and 3, we observe more scatter (especially for the former), which is likely introduced by the scatter between simulation and parameterized values of $R_\text{t}$, but also due to the inherent complexity in this transitioning state.

\begin{figure}
\begin{center}
\includegraphics[width=0.4625\textwidth]{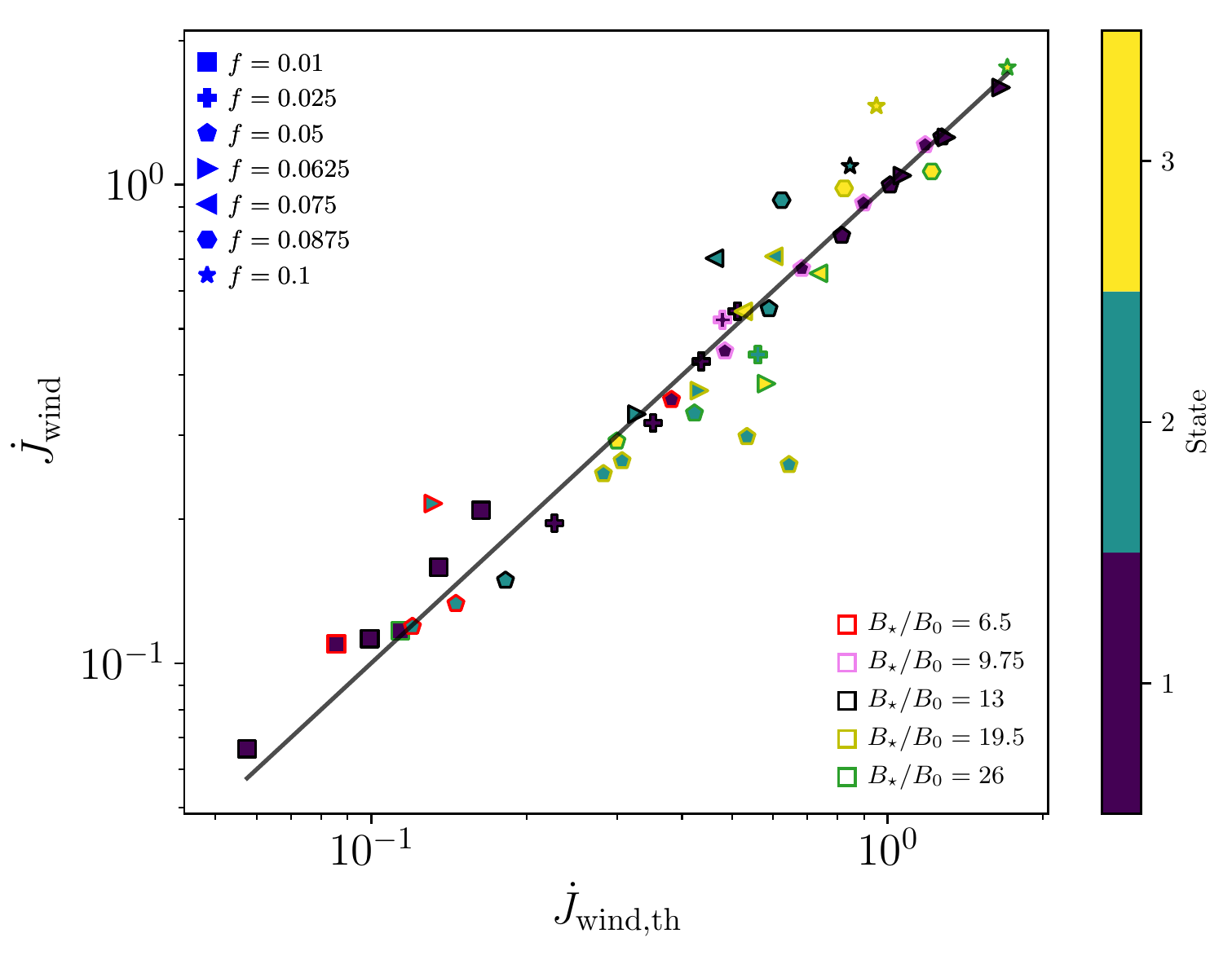}
\caption{Stellar wind torque, $\dot{J}_\text{wind}$, as a function of the parameterization in Equation~(\ref{eq:SW_torque_param}) for all simulations. Black line shows $y=x$, indicating goodness of fit. \label{fig:Jdot_wind}}
\end{center}
\end{figure}

By combining Equations~(\ref{eq:torque_wind}),~(\ref{eq:Ra_Y_wind}), and~(\ref{eq:phi_star_param}), we write the following parmeterization for the stellar wind:

\begin{align}\label{eq:SW_torque_param}
\begin{split}
\dot{J}_\text{wind} = {}& K_{\text{A},1}^2 K_{\Phi,1}^{4m_\text{A}} \dot{M}_\text{wind} (GM_\star R_\star)^{1/2} f^{1+4m_{\Phi,2}m_\text{A}} \left(\frac{R_\text{t}}{R_\star}\right)^{4m_{\Phi,1}m_\text{A}} \\
& \times \left\{\frac{\Upsilon_\star }{[1 + (f/K_{\text{A},2})^2]^{1/2}}\right\}^{2 m_\text{A}}.
\end{split}
\end{align}
In Figure~\ref{fig:Jdot_wind}, we plot $\dot{J}_\text{wind}$ as a function of the parameterization expressed in Equation~(\ref{eq:SW_torque_param}), using values of $R_\text{t}$ determined by the parameterization expressed in Equation~(\ref{eq:Rt}). For State 1, we see good correspondence between simulation values of $\dot{J}_\text{wind}$ and our parameterization. For States 2 and 3, we again observe more scatter, which is predominantly due to the scatter associated with our parameterization of $\Phi_\text{wind}/\Phi_\star$ (i.e., $R_\text{t}$).

\begin{deluxetable}{cccc}

\tablecaption{Best-fit Dimensionless Parameters from Scaling Laws in Section~\ref{sec:torque_formulation}.\label{tab:dimensionless_params}}

\tablehead{Formulation & Parameter & Value & Equations}

\startdata
$R_\text{t}/R_\star$ & $K_{\text{t},1}$ & 0.772 & (\ref{eq:Rt}) \\
 & $m_{\text{t},1}$ & 0.311 &  \\
  & $K_{\text{t},2}$ & 1.36 &  \\
   & $m_{\text{t},2}$ & 0.0597 &  \\
   & $m_{\text{t},3}$ & -0.261 &  \\
\hline
$\dot{J}_\text{SDI}$ & $K_{\text{SDI},1}$ & 0.909 & (\ref{eq:SDI_torque}) \\
 & $K_{\text{SDI},2}$ & 0.0171 &  \\
\hline
$\langle r_\text{A} \rangle /R_\star$ & $K_\text{A,1}$ & 0.954 & (\ref{eq:Ra_Y_wind}) \\
 & $K_\text{A,2}$ & 0.0284 & \\
 & $m_\text{A}$ & 0.394 & \\
\hline
$\Phi_\text{wind}/\Phi_\star$ & $K_{\Phi,1}$ & 1.62 & (\ref{eq:phi_star_param}) \\
 & $m_{\Phi,1}$ & -1.25 & \\
 & $m_{\Phi,2}$ & 0.184 & \\
% \hline
% $\dot{M}_\text{acc}$ & $K_M$ & 0.00221 & (\ref{eq:Mdot_acc_param}) \\
\enddata
\end{deluxetable}

\section{Discussion and Conclusions}\label{sec:dis_conc}

In this work, we investigate how the net stellar torque is influenced by the relative position of the truncation and corotation radii. We perform 47 2.5D axisymmetric MHD SDI simulations, changing the mass-accretion rate (via the initial accretion disk density, $\rho_{\text{d},\star}$), the stellar surface polar magnetic field strength, $B_\star$, and the stellar rotation rate (via the stellar break-up fraction, $f$). Our simulations can be categorized into three States: State 1 when $R_\text{t}<0.433 R_\text{co}$, State 2 when $0.433 R_\text{co}<R_\text{t}<R_\text{co}$, and State 3 when $R_\text{t}>R_\text{co}$. State 1 is representative of the ``steady accretion" regime, where disk material steadily accretes onto the stellar surface. In this State, our simulations are in a net spin-up configuration. State 3 is representative of the ``propeller" regime, where disk material can gain angular momentum and be ejected by the centrifugal barrier. All simulations in this State are in a net spin-down configuration. State 2 is the transition between these two contrasting configurations, where it is possible to either be in a spin-up or spin-down configuration. 

In this study, we are motivated to investigate angular momentum transport between the star and its disk during the PMS phase, for both spin-up and spin-down SDI systems. We produce an improved formulation for the sum of the accretion and ME torque (i.e., the SDI torque), and the stellar wind torque, allowing for the prediction of the net stellar torque a priori, and for the investigation of PMS rotational evolution using 1D stellar evolution codes. In Appendix~\ref{sec:stellar_evo_code}, we rewrite the guide for implementing our torque formulation in such codes (first written in Paper~\citetalias{Ireland:2021db}, Appendix~B), in order to account for the changes in the formulation since our previous work.

An important question is whether our simulations in a net spin-down configuration produce torques that are strong enough to noticeably spin-down the star on the order of a few million years. The characteristic spin-up/down timescale can be defined as

\begin{equation}\label{eq:t_SDI}
t_\text{SDI} = \frac{J_\star}{\dot{J}_\star},
\end{equation}
where $J_\star = k^2 M_\star R_\star^2 \Omega_\star$ is the stellar angular momentum ($k^2$ is the mean radius of gyration), and $\dot{J}_\star$ is the stellar torque. It is likely that these torques will play an important role in the angular momentum balance during the first few millions of years. Using the normalizations $R_\star = 2 R_\odot$, $M_\star = 0.5 M_\odot$, $\rho_\star = 10^{-12}$ g cm$^{-3}$, $P_\star = 7.44$ days (corresponding to $f=0.0625$), and $k^2 \approx 0.2$ \citep[see, e.g.,][]{1996MNRAS.280..458A}, an adimensional value of the net stellar torque equal to one in Figures~\ref{fig:Jdot_SDI_v_Jdot_SDI_theory} and~\ref{fig:Jdot_wind} corresponds to a spin-up/down timescale $t_\text{SDI} = 0.929$ Myr.\footnote{Calculation of $t_\text{SDI}$ in this section considers the stellar wind torque (see Section~\ref{sec:disc_wind} for further discussion on the impact of the stellar wind on the net stellar torque, especially in States 2 and 3).} The mean characteristic spin-down timescale for our State 3 simulations is $t_\text{SDI} \approx 0.86$ Myr. Using these normalizations give a magnetic field strength of $1.5 \lesssim B_\star \lesssim 2$ kG, a mass-accretion rate of $7 \times 10^{-12} \lesssim \lvert\dot{M}_\text{acc}\rvert \lesssim 4\times 10^{-10}$ $M_\odot$ yr$^{-1}$, and a mass-accretion rate of the disk at $R_\text{out}$ of $10^{-9} \lesssim \lvert\dot{M}_{\text{disk},R_\text{out}} \rvert \lesssim 10^{-8}$ $M_\odot$ yr$^{-1}$, for our State 3 simulations. These $\dot{M}_\text{acc}$ are low for T Tauri stars, but indeed this is the propeller regime, and $\dot{M}_{\text{disk},R_\text{out}}$ is in the right range.

It would be useful if future observations could distinguish between the States investigated in this study. The fact that State 1 spins up the star and drives corotation inward towards $R_\text{t}$, and vice versa for State 3, suggests that stars are likely driven to State 2 as their equilibrium state. In a hypothetical setting, where everything is fixed except for $\Omega_\star$, we would assume that State 2 is reached after $t \sim t_\text{SDI}$. Whether or not we would be able to observe stars in this State heavily depends on the variability of the system. On timescales of the order of $t_\text{SDI}$, global properties, such as the mass-accretion rate, rotation rate, and magnetic field strength, all evolve with time, making State 2 a moving target.  Furthermore, it is likely that some properties, e.g., the mass accretion rate \citep[see, e.g.,][]{10.1093/mnras/stw2977,10.1093/mnras/stz3398,Venuti_2021}, also vary on timescales much shorter than $t_\text{SDI}$. Thus, even if stars were to exist in State 2 ``on average", it is possible that they fluctuate between different states on the timescale of this variability.

{In addition, PMS stars are undergoing gravitational contraction; thus, if angular momentum was to be conserved, it could be expected that the star would spin up toward State 3. Hence, spin-equilibrium could be achieved in a State 3 configuration, rather than State~2, depending on how strong the gravitational contraction is. Using stellar-structure models \citep{1998A&A...337..403B}, we computed the timescale for spin-up, due to stellar contraction, equal to $2R_\star/(dR_\star/dt)$ \citep[see, e.g.,][]{Pantolmos2020}. For example, at an age of 1 and 10 Myr, a $0.5 M_\odot$ star has a contractional spin-up timescale of 6.2 and 58 Myr, respectively. For a $1 M_\odot$ star, the same timescales are 6.7 and 86 Myr, respectively. Therefore, for stars in this mass and age range, and with the normalizations given above, our simulations in State 3 should represent stars that are spinning down.}

\subsection{Truncation Radius}\label{sec:disc_Rt}

Our parameterization of the truncation radius is sensitive to its relative location to the corotation radius. In State 1, we can predict $R_\text{t}$ as a function of the ``disk magnetization" parameter, $\Upsilon_\text{acc}\propto B_\star^2 / \lvert \dot{M}_\text{acc} \rvert$, with a power-law index $m_{\text{t},1}=0.311$. This value is slightly lower than that found in Paper \citetalias{Ireland:2021db} ($0.340$). This is likely because of modifications made to the criteria that defines $R_\text{t}$. However, deviations in this value are likely within the systematic uncertainties of these simulations. This is steeper than the analytical value ($2/7$), as the accretion disk perturbs the magnetosphere.

In State 2 and 3, where accretion is increasingly inhibited by the ejection of material at the centrifugal barrier, $R_\text{t}$ becomes less sensitive to $\Upsilon_\text{acc}$ (with a power-law index $m_{\text{t},2} = 0.0597$) and is inversely proportional to $f$ ($m_{\text{t},3} = -0.261$). For State 3, this suggests that $R_\text{t}$ will be in close proximity to $R_\text{co}$ ($1.01 \leq R_\text{t}/R_\text{co} \leq 1.16$), and that extremely high magnetic field strengths (or low mass-accretion rates) would likely be required to noticeably push $R_\text{t}$ further outward.

\subsection{SDI Torque}\label{sec:disc_SDI}

For the SDI torque, i.e., the sum of the accretion and ME torque, we propose a trifurcated formulation that accounts for each State investigated in this study. In State 1, the inner disk is rotating at a roughly constant fraction of Keplerian speed, and the MEs approximately extract a constant fraction of the angular momentum from accretion, as discussed in Paper~\citetalias{Ireland:2021db}\footnote{This measure of sub-Keplerianity, i.e., $\Omega (R_\text{t}) / \Omega_\text{K} (R_\text{t})$, ignores the small $R_\text{t}$ dependence illustrated in Equation~(26) of Paper~\citetalias{Ireland:2021db}, for simplicity.}. We find the SDI torque to be $\approx 91 \%$ of the analytic accretion torque for this State, i.e., $\dot{J}_\text{SDI} = 0.909 \dot{M}_\text{acc} (GM_\star R_\text{t})^{1/2}$. Therefore, the SDI torque formulation for this State suggests that an approximately constant fraction of the angular momentum extracted by the MEs from the disk falls onto the stellar surface (further spinning up the star), with the remainder being ejected out of the domain.

In State 2, the ME torque can produce either a spin-up or spin-down torque, depending on the relative positions of $R_\text{t}$ and $R_\text{co}$; in our study, the MEs spin down the star when $R_\text{t} \gtrsim 0.6 R_\text{co}$. In State 3, the ME torque produces a strong spin-down torque contribution. In addition, the (time-average) accretion torque becomes increasingly negligible as $R_\text{t}$ approaches and exceeds $R_\text{co}$, as a majority of disk material is ejected by the centrifugal barrier. As a result, we find that a net stellar spin-down configuration is possible for our parameter regime when $R_\text{t} \gtrsim 0.7 R_\text{co}$.
In State 3, due to negligible accretion torque, we find that the SDI torque is solely a function of the analytical magnetic torque outside of $R_\text{t}$, i.e., $\dot{J}_\text{SDI} \approx \dot{J}_{\text{ME},\star} = 0.0171 B_\star^2 R_\star^6 / R_\text{t}^3$. State 2 uses an additive function of the SDI torque parameterizations used in States 1 and 3 (replacing $R_\text{t}$ with $R_\text{co}$ in the latter). We anticipate that the SDI torque in this transitional State is more complex than the formulation adopted in this study. However, for simplicity, our additive function is preferred, considering the goodness of fit achieved, but also the fact that we find a continuous function in the transition between States.

The proposed SDI torque formulation is an improvement on the accretion and ME torque formulations outlined in Paper~\citetalias{Ireland:2021db}, not only because of the simplification of combining these torque contributions into one prescription, but also due to its compatibility with both spin-up and spin-down regimes. In Paper~\citetalias{Ireland:2021db}, the ratio of the actual and analytical torque, i.e., the ``Keplerianity" of the disk at $R_\text{t}$, is parameterized only for State 1 (spin-up) cases. Furthermore, the ME torque is a function of the field twist at the disk surface at $R_\text{t}$, which itself is parameterized only for State 1 (spin-up) cases, whilst the size of the magnetically-connected region is assumed to simply scale with $R_\text{t}$. Therefore, adopting our updated formulation gives a more precise answer than extrapolating our formulation in Paper~\citetalias{Ireland:2021db} when investigating the spin-down regime.

\subsection{Stellar Wind Torque}\label{sec:disc_wind}

The stellar wind torque formulation is determined through a parameterization of the effective Alfvén radius, based on the open flux formulation in \citet{2015ApJ...798..116R}. Our simulations follow an identical $\langle r_\text{A} \rangle$-$\Upsilon_\text{wind}$-$f$ relationship (Equation~(\ref{eq:SW_torque_param})) to that in Paper~\citetalias{Ireland:2021db}, with a power-law index $m_\text{A}=0.394$. We note that the best-fit parameters for this prescription vary slightly from those in Paper~\citetalias{Ireland:2021db}, due to the extension of our parameter regime. However, deviations are likely within the systematic uncertainties of these simulations.

To compute the stellar wind torque a priori, we parameterize the fraction of open flux in an identical way to the formulation described in Paper~\citetalias{Ireland:2021db}. In general, $\Phi_\text{wind}/\Phi_\star$ inversely scales with $R_\text{t}$: the geometry of the stellar wind region in the SDI setup is primarily influenced by the location of the inner disk, as this determines the outermost closed field loop and hence the opening angle of the wind. However, star-disk differential rotation in the SDI setup opens magnetic surfaces at higher latitudes than where the field connects the stellar surface and $R_\text{t}$ \citep[see, e.g.,][]{2002ApJ...565.1205U}, which we interpret as an additional $f$ dependence. The sensitivity to $f$ has noticeably increased compared to our prior work, with the introduction of simulations outside of the ``steady accretion" regime, giving a power-law index of $m_{\Phi,2}=0.184$ (compared to 0.0614 in Paper~\citetalias{Ireland:2021db}). In State 1, $\Phi_\text{wind}/\Phi_\star \propto B_\star^{-0.828} |\dot{M}_\text{acc}|^{0.414} f^{0.184}$, demonstrating that the fraction of open flux predominantly increases as the magnetospheric pressure decreases and/or as the dynamical disk pressure increases (both allowing the disk to open more field lines to the wind). In States 2 and 3, $\Phi_\text{wind}/\Phi_\star \propto B_\star^{-0.149} |\dot{M}_\text{acc}|^{0.0746} f^{0.510}$, suggesting that the fraction of open flux predominantly increases with stellar rotation rate (i.e., decreases with $R_\text{co}$).

{As the accretion torque becomes negligible in the transition from State 2 to 3}, the stellar wind torque dominates or becomes comparable to the stellar ME torque. We are still unsure if these higher stellar wind mass-loss rates are realistic for these stars, as well as what drives these winds. The outputted mass-loss rates should not be taken as a true prediction of these simulations, as they are solely a result of unconstrained parameters, such as the coronal density, temperature, and adiabatic index. If the stellar winds were assumed to be negligible in these simulations, the net stellar torque would be shifted closer towards a spin-up configuration, i.e., the transition between spin-up and spin-down would occur at a different $\dot{M}_\text{acc}$. However, due to the strong spin-down torque from the MEs, a spin-down configuration could still be possible in this instance. {We do not perform any simulations without a stellar wind component to investigate this; however, our suggestions are justified as our analysis and formulation for the SDI torque is independent of the stellar wind, i.e., $\dot{M}_\text{wind}$.}

\acknowledgments{LGI and SPM acknowledge support from the European Research Council (ERC) under the European Union's Horizon 2020 research and innovation program (grant agreement No 682393; \textit{AWESoMeStars}: Accretion, Winds, and Evolution of Spins and Magnetism of Stars; \url{http://empslocal.ex.ac.uk/AWESoMeStars}). CZ acknowledges support from the European Research Council (ERC) under the European Union's Horizon 2020 research and innovation program (grant agreement No 742095; SPIDI: Star-Planets-Inner Disk-Interactions; \url{http://spidi-eu.org}).

The authors would like to acknowledge the use of the University of Exeter High-Performance Computing (HPC) facility in carrying out this work.

We thank Andrea Mignone and others for the development and maintenance of the PLUTO code. Figures within this work are produced using the Python package Matplotlib \citep{2007CSE.....9...90H}.}

\software{Matplotlib \citep{2007CSE.....9...90H}, PLUTO \citep{0067-0049-170-1-228,2012ApJS..198....7M}.}

\appendix

\section{Using our Formulation to Calculate Net Torque}\label{sec:stellar_evo_code}

For clarity, here is a simple guide to using the torque formulation derived in this study.

\begin{enumerate}
	\item{Specify the following input parameters: stellar mass ($M_\star$), stellar radius ($R_\star$), surface polar magnetic field strength ($B_\star$), stellar break-up fraction ($f$), mass-accretion rate ($\dot{M}_\text{acc}$), and stellar wind mass-loss rate ($\dot{M}_\text{wind}$).}
	\item{Calculate the truncation radius ($R_\text{t}$), using Equation~(\ref{eq:Rt}).}
	\item{Calculate the SDI torque ($\dot{J}_\text{SDI} = \dot{J}_\text{acc} + \dot{J}_{\text{ME},\star}$) and stellar wind torque ($\dot{J}_\text{wind}$) using Equations~(\ref{eq:SDI_torque}) and~(\ref{eq:SW_torque_param}), respectively.}
	\item{Sum each torque contribution to calculate the net stellar torque ($\dot{J}_\star = \dot{J}_\text{SDI} + \dot{J}_\text{wind}$).}
\end{enumerate}

The simulations in this study are performed over several dynamical timescales. However, the evolutionary timescales for such stars are over several million years. To predict the rotational evolution of these stars, the net stellar torque can be predicted for a given set of global properties, which can be evolved with respect to time using a stellar evolution code.
The derived torque formulation has not been tested outside of the parameter regime investigated in this study. Therefore, we recommend that this formulation is not used for conditions that give a truncation radius outside of $R_\star \leq R_\text{t} \leq 1.16 R_\text{co}$.

\bibliography{papers}

\end{document}